\documentclass[iop,apj,revtex4]{emulateapj}
\newcommand{\megasaura}{M\textsc{eg}a\textsc{S}a\textsc{ura}}
\newcommand{\megasauralong}{The Magellan Evolution of Galaxies Spectroscopic and Ultraviolet Reference Atlas}
\newcommand{\rcsohthree}{RCS-GA~032727$-$132609}

\newcommand{\sohohthreethree}{SGAS~J003341.5$+$024217}
\newcommand{\sohoneoheight}{SGAS~J010842.2$+$062444}
\newcommand{\sohninehundred}{SGAS~J090003.3$+$223408}

\newcommand{\sfifteentwentyseven}{SGAS~J152745.1$+$065219}

\newcommand{\shapenorm}{shape-normalized}
\newcommand{\pivot}{$\lambda_{pivot}$-normalized}

\newcommand{\snine}{S99-normalized}
\newcommand{\etal}{et~al.}
\newcommand\revisedApJ{}   
\newcommand{\Msun}{M$_{\odot}$}

\newcommand{\kms}{\hbox{km~s$^{-1}$}}

\newcommand{\peryr}{\hbox{yr$^{-1}$}}
\newcommand{\ciii}{[C~III]~1907, C~III]~1909~\AA}
\newcommand{\mgii}{Mg~II~2796, 2803~\AA}
\newcommand{\civ}{C~IV 1548, 1551~\AA}

\newcommand{\ArII}{\hbox{{\rm Ar}\kern 0.1em{\sc ii}}}
\newcommand{\ArIII}{\hbox{{\rm Ar}\kern 0.1em{\sc iii}}}
\newcommand{\CIV}{\hbox{{\rm C}\kern 0.1em{\sc iv}}}
\newcommand{\HI}{\hbox{{\rm H}\kern 0.1em{\sc i}}}
\newcommand{\HII}{\hbox{{\rm H}\kern 0.1em{\sc ii}}}
\newcommand{\HeI}{\hbox{{\rm He}\kern 0.1em{\sc i}}}
\newcommand{\HeII}{\hbox{{\rm He}\kern 0.1em{\sc ii}}}
\newcommand{\NII}{\hbox{{\rm N}\kern 0.1em{\sc ii}}}
\newcommand{\OI}{\hbox{{\rm O}\kern 0.1em{\sc i}}}
\newcommand{\OII}{\hbox{{\rm O}\kern 0.1em{\sc ii}}}
\newcommand{\OIII}{\hbox{{\rm O}\kern 0.1em{\sc iii}}}
\newcommand{\OIIlong}{{\rm O}\kern 0.1em{\sc ii}~$\lambda 3727$} 
\newcommand{\FeII}{\hbox{{\rm Fe}\kern 0.1em{\sc ii}}}
\newcommand{\NeII}{\hbox{{\rm Ne}\kern 0.1em{\sc ii}}}
\newcommand{\NeIII}{\hbox{{\rm Ne}\kern 0.1em{\sc iii}}}
\newcommand{\NeV}{\hbox{{\rm Ne}\kern 0.1em{\sc v}}}
\newcommand{\SII}{\hbox{{\rm S}\kern 0.1em{\sc ii}}}
\newcommand{\SIII}{\hbox{{\rm S}\kern 0.1em{\sc iii}}}
\newcommand{\SIV}{\hbox{{\rm S}\kern 0.1em{\sc iv}}}
\newcommand{\SiIV}{\hbox{{\rm Si}\kern 0.1em{\sc iv}}}
\newcommand{\MgII}{\hbox{{\rm Mg}\kern 0.1em{\sc ii}}}
\newcommand{\Halpha}{\hbox{{\rm H}\kern 0.1em$\alpha$}}
\newcommand{\Hbeta}{\hbox{{\rm H}\kern 0.1em$\beta$}}
\newcommand{\Heopta}{\hbox{{\rm He}\kern 0.1em{\sc i}}~$6678$}
\newcommand{\Heoptb}{\hbox{{\rm He}\kern 0.1em{\sc i}}~$5876$}
\newcommand{\Heoptc}{\hbox{{\rm He}\kern 0.1em{\sc i}}~$4471$}
\newcommand{\Brgam}{\hbox{{\rm Br}\kern 0.1em$\gamma$}}
\newcommand{\Brten}{\hbox{{\rm Br}\kern 0.1em$10$}}
\newcommand{\Breleven}{\hbox{{\rm Br}\kern 0.1em$11$}}
\newcommand{\HeIh}{\hbox{{\rm He}\kern 0.1em{\sc i}}~$1.7$~{\micron}}
\newcommand{\HeIk}{\hbox{{\rm He}\kern 0.1em{\sc i}}~$2.06$~{\micron}}
\newcommand{\squishlist}{
   \begin{list}{$\bullet$}
    { \setlength{\itemsep}{0pt}      \setlength{\parsep}{1pt}
      \setlength{\topsep}{3pt}       \setlength{\partopsep}{0pt}
      \setlength{\leftmargin}{1.5em} \setlength{\labelwidth}{1em}
      \setlength{\labelsep}{0.5em} } }
\newcommand{\squishend}{
    \end{list}  } 
\usepackage{pdfpages, subfloat}
\newcommand{\nstacked}{$N=14$}   
\newcommand{\samplesize}{$N=15$}  
\newcommand{\vmax}{$v_{max}$}
\newcommand{\vmaxblue}{$v_{max, blue}$}
\newcommand{\vmaxred}{$v_{max, red}$}
\newcommand{\vmaxabs}{$|v_{max}|$}
\newcommand{\vmean}{$v_{mean}$}

\newcommand{\coshst}{COS/\textit{HST}}
\newcommand{\avgR}{$R=3300$}

\slugcomment{Submitted to ApJ, 27 Jun.\ 2017.  Resubmitted 18 Oct.\ 2017}

\shorttitle{Stacked rest-frame UV spectrum at z$\sim$2}
\shortauthors{Rigby \etal}
\begin{document}

\title{The Magellan Evolution of Galaxies Spectroscopic and 
Ultraviolet Reference Atlas (\megasaura) II:
Stacked Spectra.}
\author{J.~R.~Rigby\altaffilmark{1}, 
M.~B.~Bayliss\altaffilmark{2}, 
J.~Chisholm\altaffilmark{3},  
R.~Bordoloi\altaffilmark{2,4}, 
K.~Sharon\altaffilmark{5}, 
M.~D.~Gladders\altaffilmark{6,7},
T.~Johnson\altaffilmark{5}, 
R.~Paterno-Mahler\altaffilmark{5},
E.~Wuyts\altaffilmark{8}, 
H.~Dahle\altaffilmark{9}, 
and A.~Acharyya\altaffilmark{10}}
\altaffiltext{1}{Observational Cosmology Lab,  Goddard Space Flight Center, 8800 Greenbelt Rd., Greenbelt, MD 20771, USA}
\altaffiltext{2}{MIT Kavli Institute for Astrophysics and Space Research, 
          77 Massachusetts Ave., Cambridge, MA 02139, USA}
\altaffiltext{3}{Observatoire de Gen\`{e}ve, Universit\`{e} de Gen\`{e}ve, 51 Ch. des Maillettes, 1290 Versoix, Switzerland}
\altaffiltext{4}{Hubble Fellow}
\altaffiltext{5}{Department of Astronomy, University of Michigan, 
          500 Church St., Ann Arbor, MI 48109, USA}
\altaffiltext{6}{Department of Astronomy \& Astrophysics, University of
           Chicago, 5640 S. Ellis Ave., Chicago, IL 60637, USA}
\altaffiltext{7}{Kavli Institute for Cosmological Physics, University of
          Chicago, 5640 South Ellis Ave., Chicago, IL 60637, USA}
\altaffiltext{8} {ArmenTeKort, Antwerp, Belgium}
\altaffiltext{9}{Institute of Theoretical Astrophysics, University of Oslo, 
              P.O. Box 1029, Blindern, NO-0315 Oslo, Norway}
\altaffiltext{10} {The Australian National University, Australia}
\email{Jane.Rigby@nasa.gov}

\begin{abstract}
We stack the rest-frame ultraviolet spectra of \nstacked\ highly magnified gravitationally 
lensed galaxies at redshifts \revisedApJ{$1.6<z<3.6$}. The resulting new composite spans 
\revisedApJ{$900< \lambda_{rest} < 3000$~\AA}, with a peak signal-to-noise ratio of 103 per spectral resolution 
element ($\sim$100~\kms).  It is the highest signal-to-noise ratio, highest spectral resolution
composite spectrum of $z\sim2$--3 galaxies yet published. 
The composite reveals numerous weak nebular emission 
lines and stellar photospheric absorption lines that can serve as new physical diagnostics, 
particularly at high redshift with the \textit{James Webb Space Telescope} (\textit{JWST}).
We report equivalent widths to aid in proposing for and interpreting \textit{JWST} spectra. 
We examine the velocity profiles of strong absorption features in the composite, 
and in a matched composite of  $z\sim0$ \coshst\ galaxy spectra.
We find remarkable similarity in the velocity profiles at $z\sim 0$ and $z\sim2$, suggesting
that similar physical processes control the outflows across cosmic time.
While the maximum outflow velocity depends strongly on ionization potential, the absorption-weighted 
mean velocity does not.  As such, the bulk of the high-ionization absorption 
traces the low-ionization gas, with an additional blueshifted absorption tail
extending to at least $-2000$~\kms .  We interpret this tail as arising from the stellar
wind and photospheres of massive stars. Starburst99 models 
are able to replicate this high-velocity absorption tail.  However, these theoretical models 
poorly reproduce several of the photospheric absorption features, indicating
that improvements are needed to match observational constraints on the massive 
stellar content of star-forming galaxies at $z \sim 2$. 
We publicly release our composite spectra. 
\end{abstract}


\section{Introduction}
The rest-frame ultraviolet spectra of galaxies are 
particularly rich in diagnostics of hot stars, the nebulae they ionize, 
and the outflowing winds they power.  
The \textit{International Ultraviolet Explorer} (\textit{IUE}), 
the \textit{Far Ultraviolet Spectroscopic Explorer} (\textit{FUSE}), 
the \revisedApJ{\textit{Hopkins Ultraviolet Telescope} (\textit{HUT})}, and 
four generations of ultraviolet spectrographs onboard \textit{The Hubble Space Telescope} 
have studied these diagnostics in nearby galaxies  
(e.g. \citealt{Kinney:1993eh, Leitherer:2011cg, Heckman:2011ju, Crowther:2016ip}).
Since most of the universe's stellar mass was formed 
at z$\sim$2--3 during an era of rampant star formation (compare to today at $z\sim0$), 
it is natural to compare the rest-frame UV  spectra of $z\sim0$ and $z\sim2$--3 galaxies.  
Do distant galaxies have the same populations of massive stars as nearby galaxies?  
Have galactic winds evolved with time?

Obtaining such diagnostic spectra for normal galaxies at redshifts of $1<z<4$ 
is normally beyond the reach of current instrumentation, and indeed, 
is a goal of future mission concepts such as a large ultraviolet/optical/infrared telescope or ``LUVOIR''
\citep{Kouveliotou:2014vc,Dalcanton:2015tz}.  
 Progress has been made by stacking low resolution spectra  (R$\sim$600--\revisedApJ{1400})
of galaxies at these redshifts, to study spectral features too faint to be detected  in individual spectra.  

\citet{Shapley:2003gd} stacked the low-signal-to-noise Keck/LRIS spectra of almost one thousand 
faint Lyman break galaxies at redshifts of $z\sim$2--3. 
\citet{Jones:2012kn} extended this analysis to $z\sim4$ by stacking the Keck/DEIMOS 
spectra of $N=81$ Lyman break galaxies.
\citet{Steidel:2010go} stacked the Keck/LRIS spectra of galaxies at $z\sim2.3$, and
\citet{Steidel:2016wu} stacked the Keck/LRIS spectra of 30 star-forming galaxies at $z=2.4$.
These latter two studies also stacked  rest-frame optical spectra from 
Keck/NIRSPEC or Keck/MOSFIRE for the same samples, enabling cross-comparison of the
rest-frame ultraviolet and optical emission lines.  
\citet{Zhu:2015kv} stacked spectra of 8620 emission-line galaxies at $0.6<z<1.2$, 
covering to longer wavelengths ($\lambda_{r} > 2200$~\AA ).
Table~\ref{tab:stacked} summarizes the basic parameters of the composites that resulted
from these studies -- redshift range, wavelength range, spectral resolving powers, and signal-to-noise ratio.

In this Paper, we stack the spectra of \nstacked\ gravitationally lensed galaxies from Project 
\megasaura : \megasauralong\  (Rigby \etal\ submitted; hereafter Paper I).  
The resulting stacked spectra represent a new moderate spectral resolution, 
rest-frame ultraviolet composite for star-forming galaxies at z$\sim$2. 
The high signal-to-noise ratio of this composite enables a detailed comparison with 
with stellar population synthesis models,  enabling us to constrain the stellar and galactic winds.

All magnitudes quoted are in the AB system.  

\section{Methods}
Below we describe the spectra that are stacked to generate the MagE \megasaura\ 
composite spectra,  the methodology for measuring the systemic redshifts, 
the methodology for normalizing the fluxes of the input spectra, and 
the methodology for stacking the spectra.  
For reasons detailed in \S\ref{sec:normalization}, 
rather than generating a single composite \megasaura\ spectrum, we create 
several different stacks, each with different methods for normalizing the input spectra. 
We describe how we stacked \coshst\ spectra of $z\sim0$ galaxies 
to make a composite spectrum representing local galaxies.  
We also describe the method of measuring outflow velocities.

\subsection{Input \megasaura\ MagE spectra}
The \megasaura\ sample and MagE spectra are described in detail in Paper~I.  
The \samplesize\ galaxies span the redshift range \revisedApJ{1.6$<$z$<$3.6}, and are among the
brightest lensed sources selected from the Sloan Digital Sky Survey;
the brightest have a g-band magnitude of $g_{AB} \la 21$. By selection, the \megasaura\ galaxies are 
skewed toward galaxies with high rest-frame UV surface brightness, i.e., vigorously star-forming 
galaxies \revisedApJ{with relatively low dust content.} 
We are compiling star formation rates and stellar masses for the whole 
sample; a subset have published physical properties, with stellar masses 
of 3--$7 \times 10^9$~\Msun, stellar ages of $\sim100$~Myr, $E(B-V)=0.10$--1.0, and star formation 
rates of 20--100~\Msun \peryr \citep{Wuyts:2012ej,Bayliss:2014ib}.
\revisedApJ{Seven of the \megasaura\ galaxies have metallicity measurements, 
which are tabulated in Table~2 of Paper~I.  
The measurements range from $<25\%~Z_{\odot}$  to $<81\%~Z_{\odot}$, 
with a median measured metallicity of $37\%~Z_{\odot}$.}

In brief, the \megasaura\ spectra are of high signal-to-noise-ratio 
and moderate spectral resolution.
They were obtained with the MagE instrument \citep{Marshall:2008bs} on the Magellan telescopes.
The spectra cover the rest-frame ultraviolet \revisedApJ{($900 \la \lambda_r \la 3000$~\AA). }
The spectral resolving powers range from R$=$2500 to R$=$4700.  
The median--quality spectrum has a signal-to-noise ratio of $SNR=21$ per resolution element 
at $\lambda_{obs} = 5000$~\AA .

\revisedApJ{The input spectra have been corrected for Milky Way reddening, 
using the $E(B-V)$ value derived from from Pan-STARRS~1 and 2MASS photometry by 
\citet{Green:2015cf}, assuming an extinction--to--reddening constant of 
$R_v = 3.1$, and the reddening curve of \citet{Cardelli:1989dp}.  
Table~1 of Paper I lists the Galactic $E(B-V)$ values that were used.}

We exclude the galaxy SGAS~J224324.2$-$093508 from the stack, since the MagE spectrum
reveals it to contain a broad-line active galactic nucleus (AGN) (see \S3.2 of Paper~I.)  
This reduces the number of galaxies stacked to \nstacked.

As detailed in Table~1 of Paper~I, 
for a few \megasaura\ galaxies we have obtained spectra of multiple distinct physical regions, 
because we placed spectroscopic slits at different positions along the highly magnified giant arcs. 
To prevent these galaxies from dominating the stack, we do not include the spectra of each 
physical region of a lensed galaxy in the stacking process.  Instead, we use the spectrum 
of the spectrum of the physical region with the highest signal-to-noise ratio. 
Since the stack is weighted by signal-to-noise ratio, such spectra would be the dominant contribution
from their input galaxy in any case.

\subsection{Systemic redshift}
Before stacking, the input spectra must be shifted to rest-frame wavelength.  
This is done using systemic redshifts from Paper~I.
For this sample we have access to both nebular line redshifts and stellar redshifts.  
 The stellar redshift was measured by fitting linear combinations of Starburst99 
\citep{Leitherer:1999jt, Leitherer:2010dz} simple stellar population models to each spectrum, 
while simultaneously fitting for reddening, following the methodology of \citet{Chisholm:2015fo}. 
\revisedApJ{These fits were made using a linear combination of 10 single-aged, fully theoretical 
Starburst99 models, and 5 different stellar continuum metallicities (0.01, 0,2, 0.4, 1.0, and 2.0~Z$_\odot$). 
The median light--weighted stellar metallicity is 0.37~Z$_\odot$, and the light--weighted stellar age is 10~Myr. 
The theoretical stellar continuum models were compiled using the WM-BASIC code \citep{Leitherer:1999jt, Leitherer:2010dz} 
with the Geneva stellar evolution tracks with high mass loss \citep{Meynet:1994tx}.  We used a Kroupa IMF, with a 
power--law index of 1.3 (2.3) for the low (high) mass slope, and a high-mass cut-off at 100~M$_\odot$.  
We fully discuss these model in an upcoming paper (Rigby et al.\ in prep.)\revisedApJ{.}
Given this method of fitting the stellar continuum,}
the stellar redshift is dominated by the myriad weak photospheric absorption features.
The nebular redshift was measured by fitting two Gaussians to the 
\ciii\ doublet, except as noted in Table~3 of Paper~I.
We noted in Paper~I that there is  
\revisedApJ{no systematic offset between the redshifts of the hot stars and the nebulae; the median offset 
and median absolute deviation are $-1 \pm 31$~\kms .} 

\revisedApJ{We experimented with the choice of systemic redshift for the stacks: nebular or stellar.}
Unsurprisingly, the nebular emission lines are narrower when the nebular redshifts were used as systemic.
\revisedApJ{The line profiles of the photospheric absorption lines near $\lambda_{rest} = 1300$~\AA\ 
are similar for the two choices of systemic redshift.  
Since the nebular redshifts are more precise, we use the nebular redshifts as the systemic redshifts
for all the stacks of the \megasaura\ spectra.}

\subsection{Normalization of input spectra}\label{sec:normalization}
Before stacking, we normalize each spectrum, using one of the following methods:
\begin{itemize}
\item Divide each input spectrum by its hand-fit spline continuum from Paper~I.  
This normalizes both the zeropoint and shape of each input spectrum, and 
prevents the stack from``ringing'' at the edges of the bandpass, due to 
small numbers statistics and a range of spectral slopes.  
The output spectrum  should have a flat spectral shape, 
with the exception of the region near and blueward of Lyman $\alpha$, where
continuum fitting is extremely challenging.
This stack is best suited for measuring the equivalent widths of faint
emission lines, since for this application the 
galaxy's intrinsic spectral shape is not important.  It can 
also be used to study absorption features from the interstellar medium (ISM). 
We call the composite spectrum that results from this method the \shapenorm\ stack.  

\item Divide each input spectrum by the median flux density within a pivot wavelength range:  
$1267 < \lambda_r < 1276$~\AA.
This region was chosen because it is centrally located in the spectra and 
contains relatively few spectral features.  
This normalization affects only the zeropoint of each input spectrum, and preserves the
spectral shape.  As a result, when spectra with very different slopes are averaged, 
the output spectrum may show ``ringing'' at wavelengths
far from the pivot wavelength where the spectra were normalized.
Since this stack preserves spectral shape, it is the stack to which we can fit 
stellar population synthesis models.
We call the composite spectrum that results from this method the \pivot\ stack. 


\item Divide each input spectrum by its best-fit model linear combination of Starburst~99 models
from Rigby \etal\ (in prep.)  This normalization affects both the zeropoint and the spectral shape.
This stack may be the best for analyzing ISM absorption profiles, since the stellar wind features
have already been removed.
We call the composite spectrum that results from this method the \snine\ stack.
\end{itemize}

\subsection{Stacking methodology}
We stack the \megasaura\ spectra of \nstacked\  gravitationally-lensed galaxies 
 as follows.  Each input spectrum is continuum-normalized
using one of the three normalization methods described above. 
We then shift it to its rest frame, using the nebular line redshift as the systematic redshift.
We then resample, using linear interpolation, onto a common output wavelength grid 
\revisedApJ{with a sampling of 0.1~\AA.}

We take steps to keep the composite spectra free of spurious features due to intervening 
absorption systems.  We identified such systems through a systematic search for 
intervening Mg~II, C~IV, and Si~IV doublets in all \megasaura\ spectra. 
In each input spectrum we mask  $\pm200$~\kms\ around the positions of all transitions from 
these intervening absorption systems. 

The input spectra, after having been continuum-normalized, de-redshifted, resampled, 
and masked of intervening absorption systems, 
are then stacked  using two methods:  
weighted average (weighted by the uncertainty spectra); 
and median (with no weighting).
For most applications, the weighted average is preferable to the median 
since it has higher signal-to-noise ratio.

\begin{figure}
\includegraphics[page=1, scale=0.65]{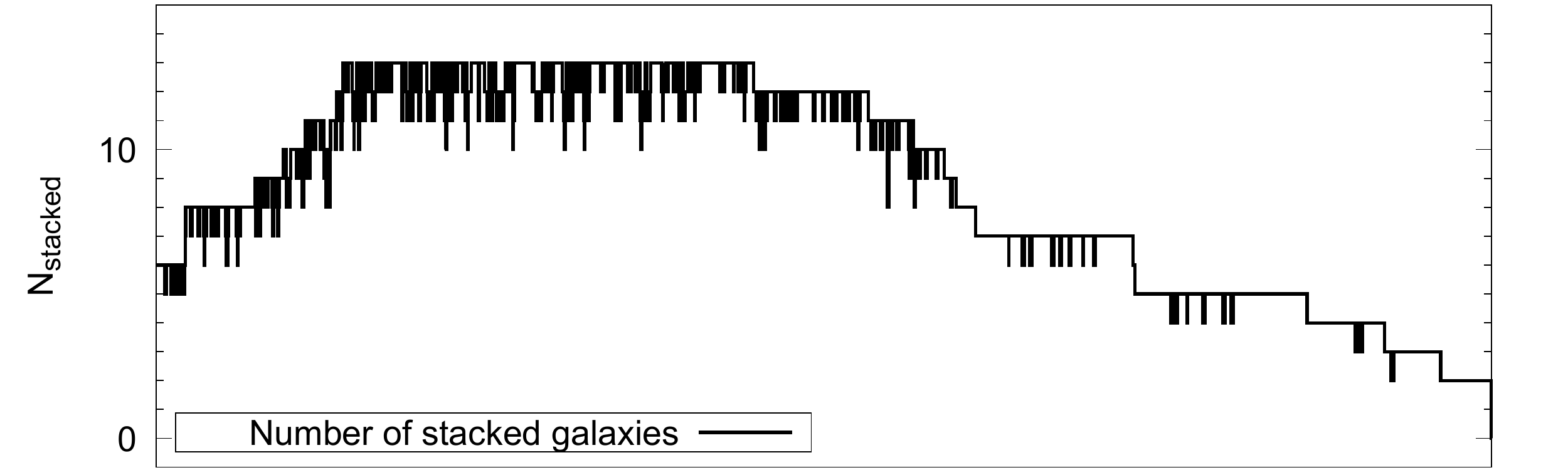} 
\includegraphics[page=3, scale=0.65]{f1.pdf} 

\includegraphics[page=4, scale=0.65]{f1.pdf} 
\figcaption{The \megasaura\ \shapenorm\  stacked  spectrum. 
The top panel shows the number of galaxies that went into the stack at each rest-frame wavelength; 
that number is affected by the observed wavelength coverage of the input spectra, 
\revisedApJ{ the location of saturated skylines, and the expected positions of intervening absorption lines.}  
The middle panel shows two independent measures of the signal-to-noise ratio, 
\revisedApJ{per resolution element, 
of the weighted average stacked spectrum:} 
the jackknife uncertainty,  and the propagation of the individual uncertainty spectra.  
The signal-to-noise ratio curves have been smoothed for readability.
The bottom panel compares two realizations of the \shapenorm\ stack: the weighted average 
\textit{(black)} to the median \textit{(grey)}.\label{fig:snr}}
\end{figure}

To understand the uncertainty in the composite spectra 
we perform jackknife tests, in which we compute the weighted average $n$ times, each with a 
different spectrum masked out.  For each stack, we compute two estimates of the per-pixel 
uncertainty:
a) the uncertainty spectrum propagated from the individual uncertainty spectra, and 
b) the jackknife uncertainty $\sigma_{jackknife}$, 
which measures the variation in the $n$ jackknife spectra: 
\begin{equation}
\sigma_{jackknife}^2 =   \frac{(n-1)}{n}   \sum\limits_{i=1}^{n} ( x_i - x_{(.)})^2  
\end{equation}
where $x_i$ is the pixel value in the \textit{i}th jackknife spectrum, and 
$x_{(.)}$ is the pixel value in the weighted average spectrum.
These two uncertainty estimates, which should be independent, are consistent, 
as Figure~\ref{fig:snr} illustrates. 
Used together, the median spectra and the jackknife uncertainty spectra can be used to 
check whether a particular spectral feature is unduly influenced by a single galaxy.

\subsection{Stacking \coshst\ spectra}
To enable an apples-to-apples comparison of the MagE \megasaura\ 
stack to the spectra of galaxies in the nearby universe, 
we stack the spectra of $z \sim 0$ galaxies with \coshst\ spectra 
\revisedApJ{using the G130−M and G160−M gratings}
from \citet{Chisholm:2016bi}.   That local sample is comprised of 
41 galaxies\footnote{We dropped one galaxy, M~83~1, from the stack due to 
difficulty fitting its continuum; this galaxy has extremely high super-solar metallicity, unlike
the \megasaura\ sample.}
 with star formation rates of 0.01 -- 137~M$_\odot$~yr$^{-1}$. 
Each galaxy has blueshifted ISM absorption lines at the 1$\sigma$ significance level. 
The stack represents the most complete sample of local star-forming galaxies 
with \coshst\ spectra that show galactic outflows.

\revisedApJ{The input COS spectra requires extra pre-processing. 
We mask the expected wavelengths of geocoronal and Milky Way emission lines.
The spectral resolution of COS is higher than that of MagE: 
for a point source it is $R=16000$, but it decreases as emission fills 
the  circular aperture of COS. In the COS spectra we are stacking, 
the Milky Way absorption lines indicate effective spectral resolutions
ranging from $R=4000$ to $R=14000$,  with a median of $R=9000$, 
due to the diffuse, extended morphology of these galaxies in the slit. 
Therefore, we resample the wavelength grid of each input COS spectrum to be 
Nyquist sampled for the spectral resolution measured from the Milky Way absorption lines,
then convolve by a Gaussian kernel to lower the spectral resolution to that of MagE (\avgR),
and then resample the wavelength array once more, to be Nyquist sampled at \avgR .   
Downgrading the effective spectral resolution in this way }
does not meaningfully change the profiles of absorption lines in the stacked spectrum, 
but it does change the per-pixel signal-to-noise ratio, and as such,
the measured maximum velocity.

Given the large number of COS spectra, rather than fit the continuum by hand as for the MagE spectra, 
we fit the continuum automatically, by masking spectral features and then convolving with a boxcar.  
Tests using the individual MagE spectra show that this automatic continuum fitting process
produces continuua that closely track our hand-fit continuua.
We then stacked the COS shape-normalized input spectra in the same way as we stacked
the MagE spectra.  The result is a shape-normalized COS stacked spectrum 
whose spectral resolution and signal-to-noise ratio closely match that of the MagE
stack (as quantified in Table~\ref{tab:stacked}.)

\subsection{Velocity measurements}
We compute two metrics of the absorption line velocity profiles:
the maximum blueshifted velocity \vmax , and 
the absorption-weighted mean velocity \vmean. 
We used the weighted-average, shape-normalized stack to measure these metrics, for both the
MagE stack and the COS stack.

We calculate the absorption-weighted mean velocity of the absorption 
line, \vmean,  within the interval from \vmaxblue\ to \vmaxred .    
We restrict \vmaxred $< 600$~\kms\ to prevent runaway fits in the case of 
doublets and other closely-spaced absorption lines (for example, for C~IV~1548.) 

\citet{Steidel:2010go} defined \vmax\ as the velocity where the blue wing of the 
absorption returns to the continuum, specifically the first pixel to return.
We adopt this definition for both \vmaxblue\ and \vmaxred,  with a caveat to the reader that 
this metric is sensitive to the signal-to-noise ratio and dispersion of the spectrum.  
This is indeed why, when we stack the $z\sim0$ \coshst\ 
spectra, we downgrade the resolution to create a stack with 
similar spectral resolution and noise properties to the MagE stack.

To gauge the uncertainty in \vmax\ and \vmean, we scale the continuum by a factor
over the range $\pm 2\%$ (as this seemed the maximum plausible error in the continuum), 
measure the mean and maximum velocities each time, and quote the mean value and standard deviation. 

\subsection{Measuring equivalent widths}
We measure the equivalent widths of emission lines in the \shapenorm\ stacked spectrum 
using the following methodology.
We fit each group of neighboring spectral lines simultaneously, with one Gaussian per feature, using 
the non-linear least-squares method as implemented in the Python tool scipy.optimize.curve\_fit.
Neighboring lines are defined as being separated by $\le 5$ spectral resolution elements. 
We use the equivalent width significance criterion of \citet{Schneider:1993hx}.

\section{Results}
\subsection{The stacked spectra}
The stacked spectra span a rest-frame wavelength range of  
\revisedApJ{$900< \lambda_{rest} < 3000$~\AA}, with 0.1~\AA\ pixels.
\revisedApJ{We calculate the median signal-to-noise ratio per resolution element in four regions
that are free of spectral lines:
$SNR=76$ for   1440--1450~\AA, 
$SNR=92$  for 1460--1470~\AA,  
$SNR=79$  1680--1700~\AA, and 
$SNR=101$  1760--1800~\AA.
}  
Figure~\ref{fig:snr} plots \revisedApJ{the stacked spectrum, 
the signal-to-noise ratio  per resolution element, and }
the number of galaxies that went into the stack at each wavelength.
The average spectral resolving power of the input spectra is \avgR .
Table~\ref{tab:stacked} compares these metrics to those of composite spectra in the literature.

\begin{figure*}
\includegraphics[width=6in]{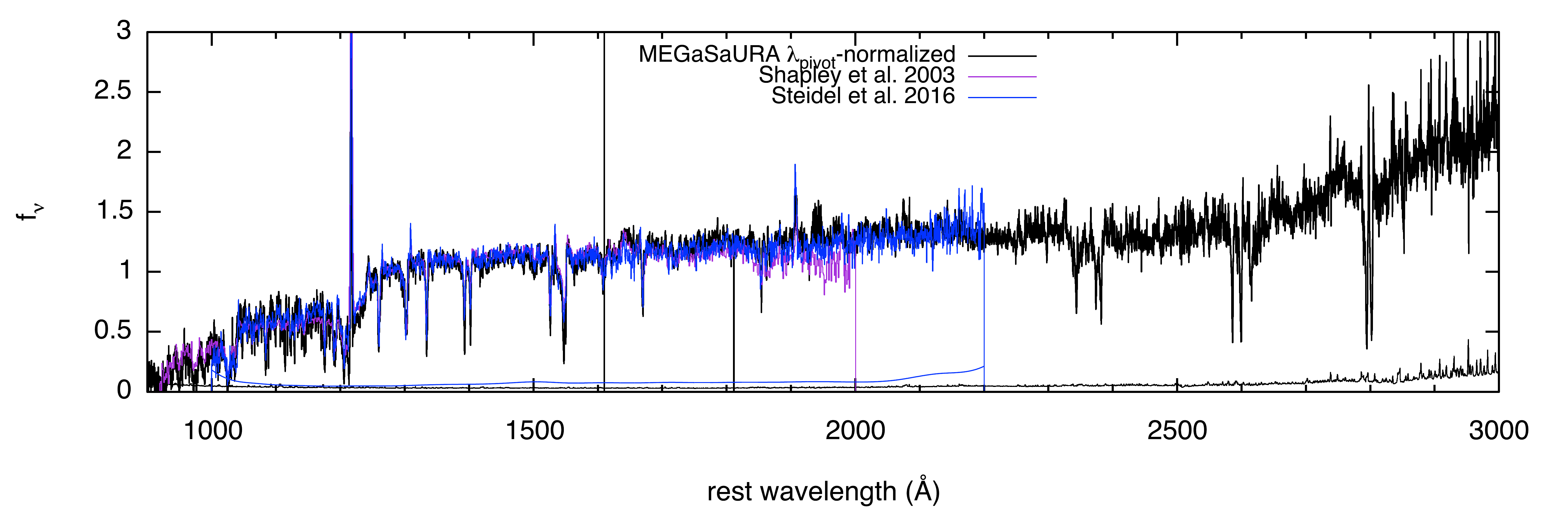} 

\includegraphics[width=3.2in]{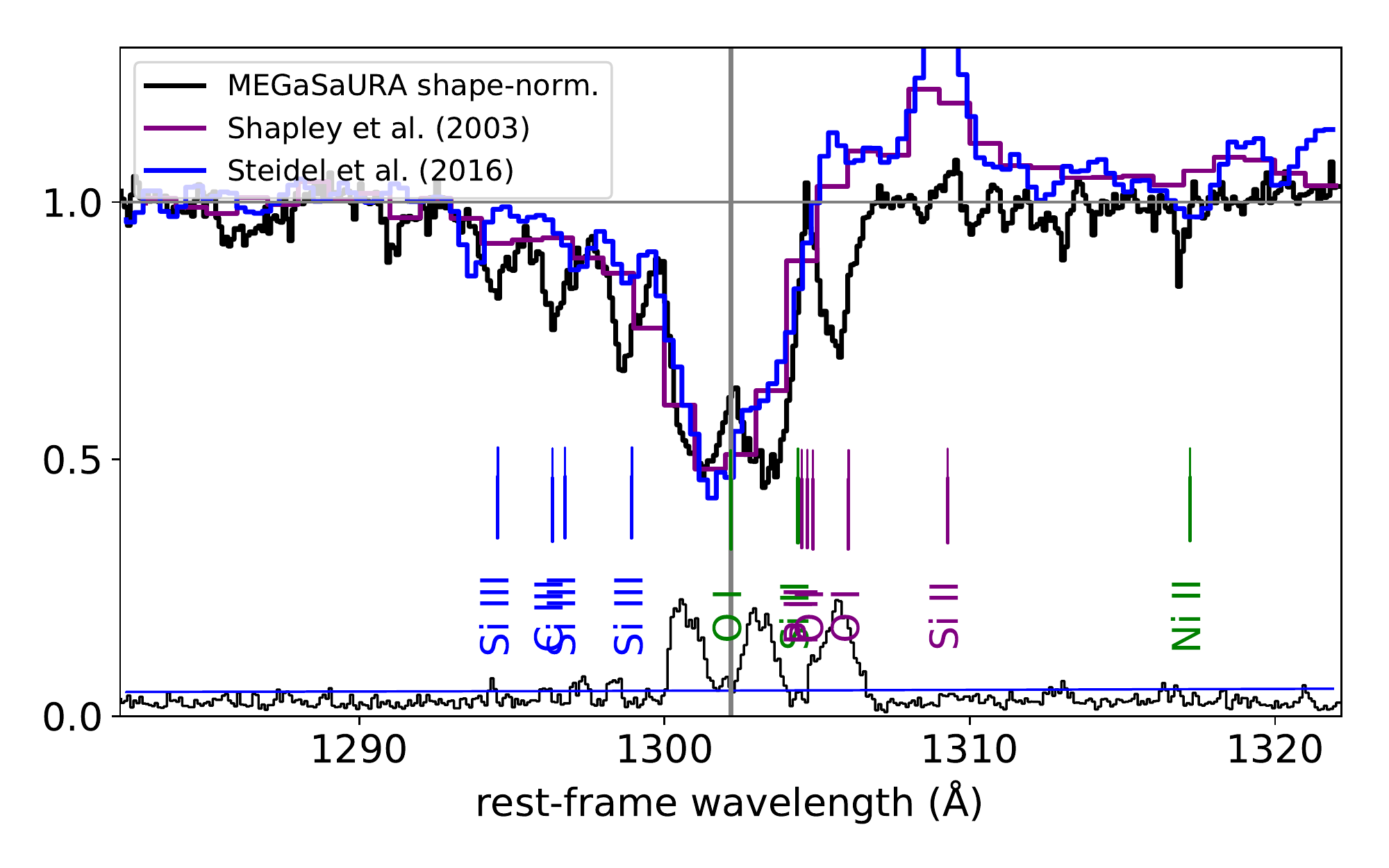} 
\includegraphics[width=3.2in]{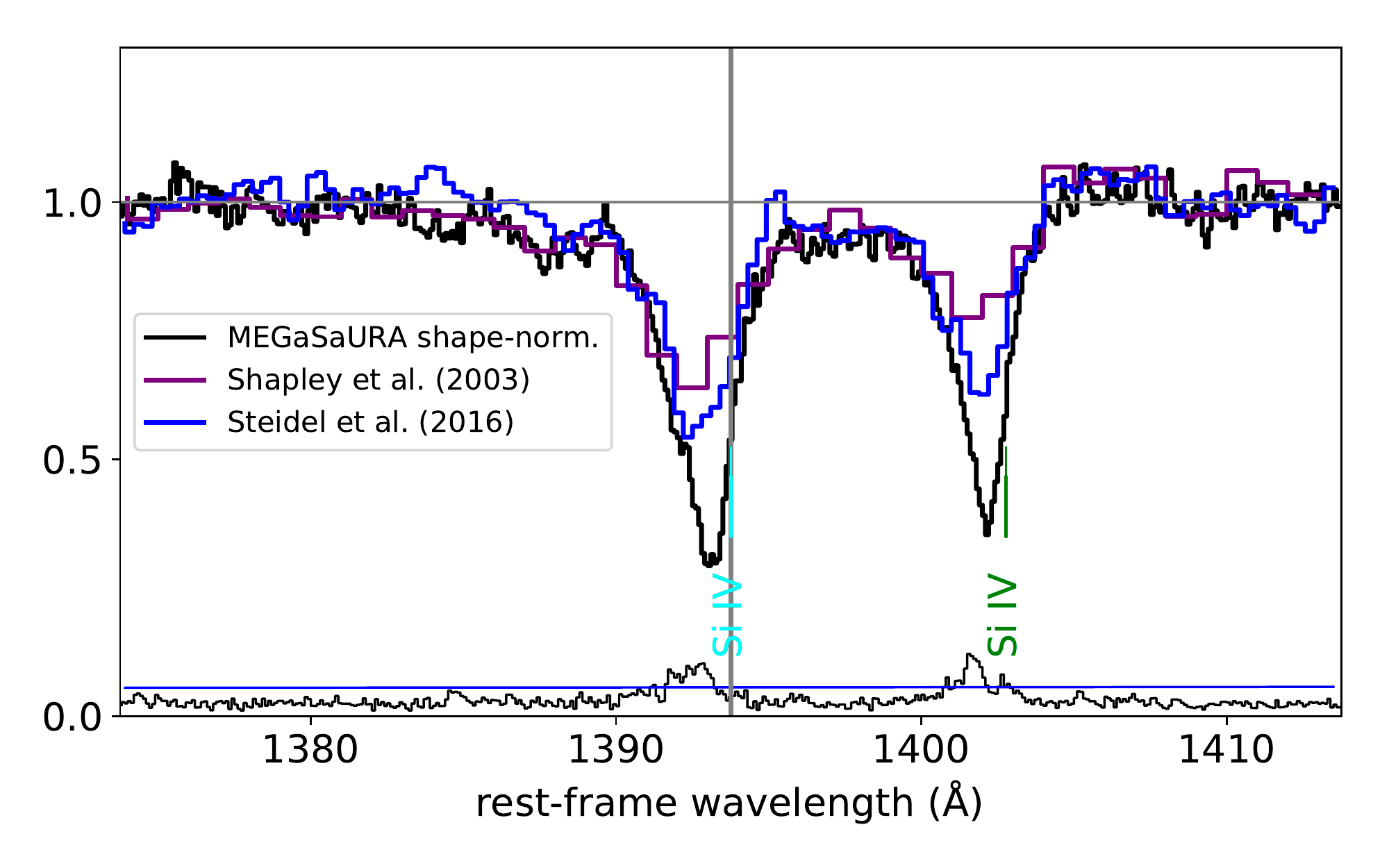} 

\includegraphics[width=3.2in]{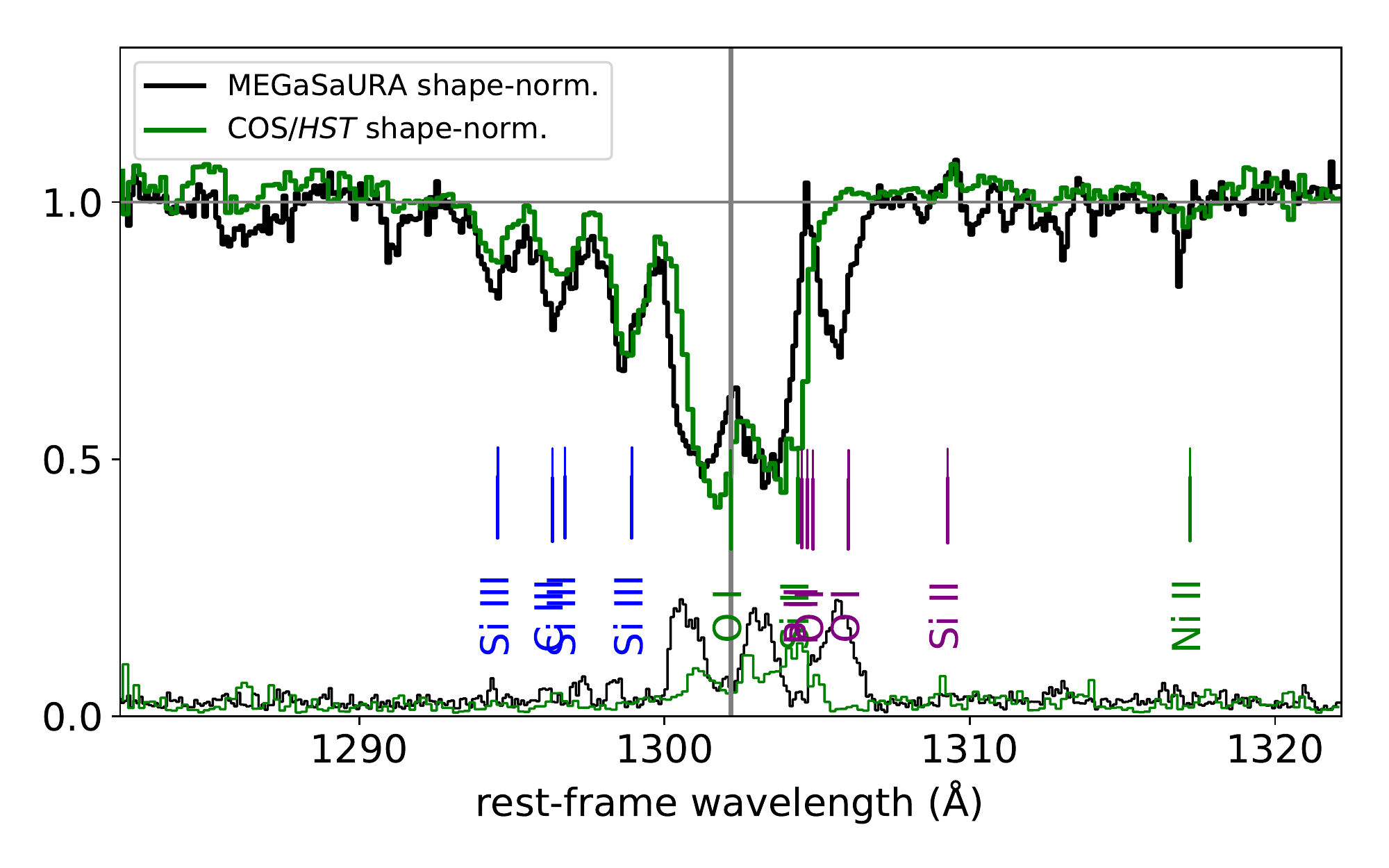}
\includegraphics[width=3.2in]{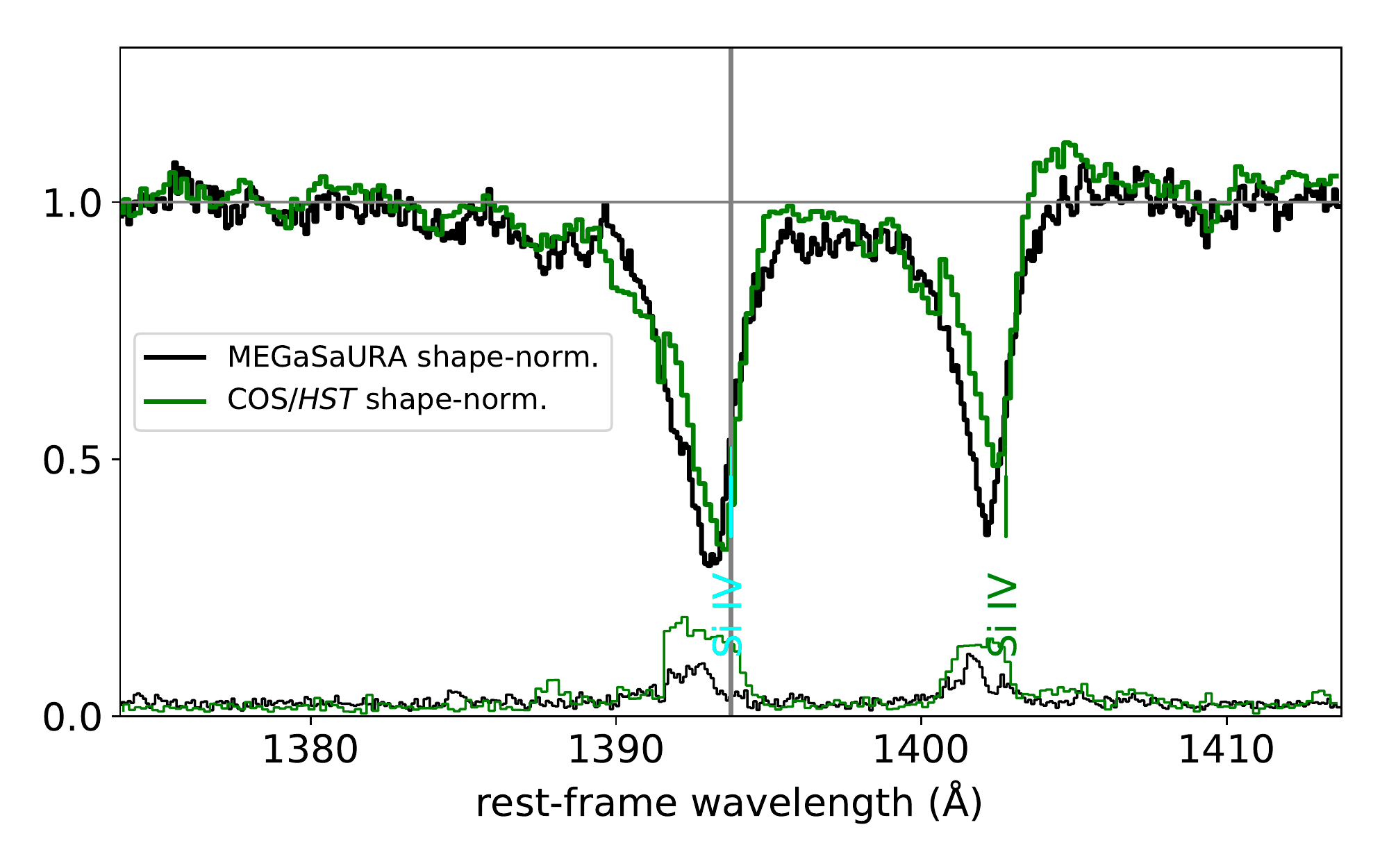} 
\figcaption{\textit{Top Panel:} Comparison of the \megasaura\ \pivot\ stacked spectrum \textit{(black steps)} to 
previously-published templates:  the composite spectra of \citet{Shapley:2003gd} \textit{(purple steps)} 
 and \cite{Steidel:2016wu} \textit{(blue steps)}.  Each plotted spectrum has been renormalized, 
by dividing the flux density by the median in the range $1267 < \lambda_r < 1276$~\AA. 
Propagated uncertainty spectra are plotted
with the same color scheme as the spectra.
\textit{\revisedApJ{Middle} Panels:}  Zoom in on the templates for two spectral regions of interest.  
The left middle panel shows the complex spectral region near 1300~\AA\ that includes multiple 
photospheric absorption lines of Si~III and C~III, 
O~I~1302 ISM absorption, and Si~II~1309 fine structure emission. 
The right middle panel shows the region near the Si~IV doublet.
Vertical ticks and labels mark features of interest, color-coded as 
photospheric absorption  \textit{(blue)}, interstellar medium (ISM) \textit{(green)}, and stellar winds \textit{(cyan)}.
\textit{Bottom Panels:}  The same spectral regions as the middle panels, but now comparing the 
megasaura\ shape-normalized stack \textit{(black steps)} to the \coshst\ $z\sim0$ stack \textit{(green steps)}. 
Compared to previous templates, the higher spectral resolution \megasaura\ stack  much more clearly detects
weak features like the photospheric absorption lines.
\label{fig:comparetolit}}
\end{figure*}

Figure~\ref{fig:comparetolit} and Table~\ref{tab:stacked}  
demonstrate that the \megasaura\ stacked spectrum has a broader rest-frame wavelength coverage, higher 
signal-to-noise ratio, and higher spectral resolution than previously published composite spectra 
made from starburst/LBG field galaxies \citep{Shapley:2003gd,Steidel:2016wu}. 
The subpanels of Figure~\ref{fig:comparetolit} 
show that, unlike previous composites, the \megasaura\ stack has the 
combination of spectral resolution and signal-to-noise ratio required to clearly detect
faint spectral features and to measure the shapes of absorption lines.  
The composite \megasaura\ spectra therefore provide the best census of rest-frame UV properties of 
starburst galaxies in the era of ``cosmic noon''. We publicly release the \megasaura\ 
composite spectra, and encourage their use as templates for understanding $z\sim2$ star-forming galaxies.  

In Figure~\ref{fig:multipanel_spectrum} we show the \shapenorm\ stack, highlighting 
a number of diagnostic lines, including transitions that arise variously in the ISM, 
nebular gas, and stellar photospheres. 

For those interested in the strengths of faint emission lines, 
we recommend using the \shapenorm\ weighted average stack.
For those interested in the overall spectral shape, 
we recommend using the \pivot\ weighted average stack.

\subsection{Detected emission lines}
Table~\ref{tab:emissionlines} reports the equivalent width measurements for 
$N=18$ rest-frame UV emission lines detected in the 
\megasaura\ \shapenorm\ stacked spectrum. 
Table~\ref{tab:emissionlines}  also reports upper limits for 
undetected emission lines that we expect to be generated in the ionized gas in starburst galaxies.
The rest-frame UV emission lines should be useful as diagnostics
of physical conditions such as the pressure of the interstellar medium (Kewley \etal\ submitted to ApJ), 
electron temperature (Nicholls \etal\ in prep.), 
ionization parameter, metallicity, and the relative abundance pattern  
(\citealt{Bayliss:2014ib}, Kewley \etal\ in prep.)
By publishing the measured equivalent widths of the composite, we hope to assist 
observers in estimating the integration times required to detect these lines at high 
redshift with \textit{JWST}.

\revisedApJ{Narrow He~II~1640~\AA\ is not detected in the stacked spectrum. 
However, a broad (FWHM $\sim 2800$~\kms) excess of emission is seen over
the continuum level --- this  may be stellar in origin.
Table~\ref{tab:emissionlines} quotes an equivalent width limit for the undetected 
narrow component, and an equivalent width measurement for the broad component.}

\subsection{Comparison of velocity profiles with low redshift}
We now consider the velocity profiles of galactic and stellar winds as seen in 
the stacked \megasaura\ spectra. The \megasaura\ stacks have the high 
signal-to-noise that is required to probe outflows at the highest velocities, where the 
absorption deficit drops to of order a few percent of the continuum, and are of sufficient 
quality to enable apples-to-apples comparisons between distant lensed starburst galaxies 
and some of the most vigorously star-forming galaxies in the nearby universe.

Figure~\ref{fig:likeheckmanfig} compares the velocity profiles of strong
absorption features in the two different stacks: 
the \shapenorm\ stacked spectrum of the \megasaura\ galaxies, 
and our stack of the $z\sim0$ galaxy COS spectra from \citet{Chisholm:2016bi}.  
The velocity profiles of the $z\sim2$ stack 
are remarkably similar to those of the $z\sim0$ stack, suggesting 
gross similarity  between the galactic winds at these two very different epochs.

Table~\ref{tab:vmax} quantifies this similarity; it tabulates 
absorption-weighted mean velocity and maximum velocity  measurements for
the \megasaura\ \shapenorm\ stack, and for the \coshst\ stack, for the major 
absorption lines in the spectra.   
The next three subsections analyze the measurements in Table~\ref{tab:vmax}.

\subsection{Galactic winds as probed by Mg~II and Fe~II}\label{sec:MgIIFeII}
The wavelength coverage of the \megasaura\ stacked spectra is sufficiently broad 
\revisedApJ{($900 < \lambda_{rest} < 3000$~\AA)} to cover a large number of transitions, 
spanning a large range of ionization potential, that can be used to characterize the 
galactic and stellar winds.  In Figures~ \ref{fig:windsb}, \ref{fig:windsa}, and \ref{fig:windsc}, 
we plot the absorption profiles for several absorption lines of interest. 

In Figure~\ref{fig:windsb}, we consider the transitions of  Mg~II~2796, Fe~II~2344, and 
Fe~II~2383.\footnote{Fe~II~2600 is unusable due to adjacent Mn~II absorption.}
These transitions have ionization potentials of 15.0 eV (Mg~II) and 16.2~eV (Fe~II), 
and are associated with spatially-resolved, large-scale outflows
in galaxies at $z\sim1$--2 \citep{Weiner:2009cf, Martin:2009fl, Rubin:2010iw, Rubin:2011bm, 
Giavalisco:2011gla, Erb:2012jc, Kornei:2013ez, Rigby:2014hq, Rubin:2014hv, Bordoloi:2014fk}.
In the \megasaura\ weighted-average \shapenorm\ stack, these 
transitions have a maximum velocity of 
\vmaxabs $=  900$~\kms\   
 (Table~\ref{tab:vmax}.)

\citet{Bordoloi:2016js} analyzed the velocity profiles of four regions within one of these galaxies, 
\rcsohthree, and found  95th percentile velocities of 460--610~\kms\ for these transitions.  
Mg~II is at the extreme red end of the \megasaura\ stack, and so only four galaxy spectra went into the stack;
seven galaxies contribute to the stack for Fe~II~2344 and Fe~II~2383.   
Knot E of \rcsohthree\ contributed to the stack for both Mg~II and Fe~II.
Despite those small numbers, the maximum velocities we measure for Mg~II and Fe~II 
confirm and extend the conclusion  of \citet{Bordoloi:2016js}: 
a substantial fraction of the Mg~II and Fe~II gas column has velocities far 
exceeding the plausible escape velocity of their galactic hosts.

\subsection{Galactic winds as probed by other low-ionization transitions}\label{sec:lowion}
We next analyze the absorption velocity profiles of other low-ionization transitions, 
plotted in Figure~\ref{fig:windsa}. 
The main transitions that trace the neutral gas are Ly$\alpha$ and O~I~1302, since they have
low ionization potentials (13.6~eV).   
However, a maximum wind velocity cannot be measured for O~I~1302 
because of nearby photospheric absorption lines, and Ly$\alpha$ is difficult 
to interpret due to resonant scattering.
Instead, we must estimate the wind velocity from
Si~II, Al~II, and C~II, which have ionization potentials of 16.3, 18.8, and 24.4~eV, 
and therefore trace a mixture of neutral and ionized gas.  
This absorption has \vmaxabs $\sim$ \revisedApJ{630--940~\kms} in the  \megasaura\ weighted-average \shapenorm\ stack. 

The maximum velocities we measure, for both the MagE and COS stacks, 
are comparable to the \vmaxabs $ = \sim 700$-800~\kms\ measured for Si~II and C~II in 
the composite stacked spectra of $z\sim2.3$ galaxies of \citet{Steidel:2010go}, and 
comparable to the value of \vmaxabs $ = 900$~\kms\ measured in  Si~II~1260 
for the  $z\sim0$ stack of \citet{Alexandroff:2015cj} (see Figure 1 of \citealt{Heckman:2015dq}).

In Figure~\ref{fig:vIP} we plot the maximum velocity (blue points) and mean velocity (red points)
versus ionization potential for the MagE and COS stacks. 
For the low-ionization  lines ($IP < 25$ eV), we find that the velocity 
does not strongly depend on ionization potential. 
Rather, the mean velocities for both samples are between \revisedApJ{$-50$ and $-300$~\kms ,}
and the maximum velocities for both samples cluster between $-500$ and $-1000$~\kms.
These results are consistent with those of \citet{Chisholm:2016bi}, who found that the overall
 strength of the transition, not the ionization potential, determines
 the measured velocity. The maximum and mean velocities  
\revisedApJ{are very similar between the MagE and COS samples for are very similar
for each low-ionization transition.}
Further, the velocity profiles are very similar for these
two very different epochs (see Figure~\ref{fig:likeheckmanfig}.)
\revisedApJ{We conclude that} the average galactic outflow does not appreciably change
 from $z\sim2$ to $z\sim0$, \revisedApJ{which suggests}  that similar physical processes
 (acceleration mechanisms, for example) establish the outflow
 profiles.

Thus, the galactic wind in the \megasaura\ stacked spectrum shows a 
maximum blueshifted velocity that approaches but does not exceed 1000~\kms. 
Having established the maximum velocity of the galactic wind as traced by low ionization lines,
we now examine the maximum velocities seen in high ionization transitions.

\subsection{Stellar winds as probed by high-ionization transitions}\label{sec:highion}
In Figure~\ref{fig:windsc} we plot the absorption profiles of transitions 
with high ionization potentials, 28.4 to 97.9~eV.
These include transitions that show classical P Cygni line profiles,
and are assumed to arise in stellar winds.  
Surprisingly weak are the emission components of N~V~1238, Si~IV~1393, and C~IV~1548.  
The weakness of this emission indicates that Wolf Rayet stars and/or the most massive O stars 
are present, but rare. 

We consider the absorption profiles of these high ionization lines.
N~V~1238 is difficult to measure given the complexity of the continuum near Ly~$\alpha$.
C~IV~1548, Si~IV~1393,  and Al~III~1854 all show similar velocity profiles:
strong absorption at velocities close to systemic,  presumably from the interstellar medium,
as well as a blueshifted absorption tail extending to
\vmaxabs $=$ \revisedApJ{1300} -- 2700~\kms.  
In C~IV, this blueshifted absorption tail is detectable in 
individual \megasaura\ spectra with high signal-to-noise ratio, namely 
\rcsohthree, \sohoneoheight, \sohohthreethree, 
\sohninehundred, the Cosmic Horseshoe, and \sfifteentwentyseven .

Al~III~1854, with an ionization potential of 28.4~eV, is variously considered 
in the literature to be either an ISM or a stellar wind line.  
In the stacked spectrum, the 
high \vmaxabs\ and shape of its high-velocity tail strongly suggest that at least a substantial 
portion of Al~III~1854 absorption has a similar origin to C~IV and Si~IV.  This is supported by 
the fact that Al~III~1854 has been observed with a P Cygni profile in certain B supergiants of 
spectral and luminosity classes: B0.7 Ia to B2.5 Ia;  B1--B3 Ia+/Iap; and BC1.5 Iab 
\citep{Walborn:1995uq}. 
Thus, the Al~III profile further underscores the large contribution to the spectra from B stars.

In Figure~\ref{fig:vIP}, we plot the maximum and mean velocities of the
high-ionization absorption lines. For both the MagE and COS samples,
there is a trend where the maximum velocities increase with increasing
ionization potential, but the absorption-weighted mean velocities remain 
nearly constant for all ionization potentials. This indicates that the 
high-ionization lines have a weak blue-velocity tail extending to more than 
$\sim 2000$~\kms, but that the majority of the absorption has similar
velocities to the low-ionization gas. This supports our hypothesis
that the blue-velocity wings arise in the hot photospheres of massive
stars, while the stronger absorption component at redder velocities
arises from a multiphase galactic outflow. Properly accounting for the
stellar winds is important when measuring the maximum outflow
velocities of high-ionization lines.

The absorption in this high-velocity tail is only 6--$10\%$ below the continuum level for 
Al~III and Si~IV, and only $15\%$ for C~IV.  As such, detecting it requires 
high signal-to-noise spectra like those of the \megasaura\ stacked spectrum.
This high-velocity absorption tail
\emph{is not seen} in the profiles of transitions with lower ionization potential, 
as quantified in \S\ref{sec:MgIIFeII} and \S\ref{sec:lowion}.
This correspondence supports our hypothesis that the blue absorption wings of the
high ionization lines are likely due to the stellar winds of massive stars.

\subsection{Stellar winds fit by stellar population synthesis}
We now consider the extent to which Starburst99 stellar population synthesis models 
can reproduce these high-velocity stellar wind features. 
Thus far, we have considered the \shapenorm\ stack, which minimizes undesirable ringing and is 
therefore ideal to study these low level features.  However, for spectral synthesis fitting we 
use the \pivot\ stack, since it preserves the spectral shape. 
We plot the low-ionization features in the \pivot\ stack in Figure~\ref{fig:s99a},
and the high-ionization features in  Figure~\ref{fig:s99c}.
For both  Figure~\ref{fig:s99a} and  Figure~\ref{fig:s99c}, we 
overplot the best-fitting linear combination of Starburst99 models. 
The Starburst99 fit decently matches the N~V, Si~IV, and C~IV emission features,
as well as the high-velocity tail of C~IV.  

As a cross-check, we examine the profiles of these stellar wind lines in the 
\snine\ stack.  Because each input spectrum was continuum-normalized by its 
Starburst99 fit before stacking, the \snine\ stack provides a different way to test whether the
Starburst99 models can match the the observed high-velocity absorption profiles.
Figure~\ref{fig:snine} shows negligible residual absorption at high velocities, 
indicating that the Starburst99 models can indeed reproduce the 
high-velocity tail of  absorption from stellar winds.

\section{Discussion and conclusions}
We have stacked the MagE/Magellan spectra of \nstacked\ gravitationally lensed galaxies 
from \revisedApJ{P}roject \megasaura\ (Paper~I),  to produce 
a new spectral composite of star-forming galaxies at redshift $z\sim2$, 
one with the requisite high signal-to-noise, moderate spectral resolution, 
and wavelength coverage  that are needed to constrain the massive star populations and 
the kinematics of outflowing gas.  
We publish electronic versions of the \megasaura\ stacked spectra
to encourage their use by the scientific community.

We detect $N=18$ rest-frame UV emission lines in the \megasaura\ stack. 
These include relatively strong lines---\ciii\ and \mgii\ with $W_r \sim 0.5 $~\AA --- 
as well as weaker lines of 
He~II, C~II, N~II, [O~II] and  O~III], Si~II and Si~III, and Fe~II.  
We publish equivalent widths for these lines, to assist in estimating 
spectroscopic exposure times for \textit{JWST}.

To enable consistent  comparison between galaxies in the distant universe
and the local universe, we stack the \coshst\ spectra of $z \sim 0$ galaxies
with outflows, with the spectral resolution downgraded to approximate that of MagE.
We publish this stack \revisedApJ{\coshst} stack as well.
A surprising result from this work is how similar are the outflow
velocity profiles in the $z \sim 2$ MagE stack and in the $z \sim 0$ \coshst\ stack. 
This similarity suggests that similar physical processes may be driving the outflows 
at both epochs.  We plan to follow up this result by examining the outflows on a 
galaxy-by-galaxy basis using the individual \megasaura\ spectra.

In the \megasaura\ composite, the absorption-weighted mean velocities are similar for ions
of low and high ionization potential, which indicates that most of the highly-ionized gas 
is part of the multi-phase outflow that is traced by the low-ionization absorption lines. 
Very similar behavior is seen in our stack of $z\sim0$ galaxy spectra from 
\coshst .  For both the $z\sim2$ MagE stack and the $z\sim0$ COS stack, 
the high ionization lines show a blue tail of absorption, 
with maximum velocity of up to $-2700$~\kms\ and \revisedApJ{$-1300$}~\kms\  
in the wings of C~IV 1548 and  Si~IV~1393, respectively for the MagE stack, 
\revisedApJ{and $-2400$~\kms\ and $-1800$~\kms\ for the COS stack.}

At velocities above $-1000$~\kms, the absorption is only 6--$15\%$ of the continuum, 
and therefore requires high signal-to-noise to be detectable. 
This high-velocity absorption tail is not seen in the profiles of transitions that trace neutral 
and low-ionization gas. 
The most plausible explanation is that this high-velocity tail is a  
stellar wind feature originating from the regions around hot stars.  Using two different 
methodologies, we confirm that linear combinations of current Starburst99 models are able to 
reproduce this blue tail of high-velocity absorption in the MagE stack.  
As improvements are made to stellar population synthesis models, they will 
need to demonstrate that they, too, can reproduce this high-velocity absorption tail.

While the Starburst99 fitting is able to reproduce the overall continuum shape of the 
\megasaura\ \pivot\ stack as well as the P~Cygni stellar wind features, it does not 
adequately reproduce some of the photospheric absorption features.
This is clearly seen in the second panel and the last panel 
of  Figure~\ref{fig:s99a}, where 
a complex of C~III and Si~III lines at $\lambda_{rest} = 1296$~\AA\ 
and a blend of C~II and N~III at $\lambda_{rest} = 1324$~\AA\
are poorly matched by the Starburst99 fitting.    
These rest-frame ultraviolet photospheric absorption lines provide an additional set of 
constraints to population synthesis models; we will examine them in more 
detail in a future paper.

The composite spectrum presented in this Paper is timely, 
as we are about to enter a golden age for rest-frame ultraviolet astronomy.  
The Near Infrared Spectrograph (NIRSpec) onboard \textit{JWST} 
will soon detect the strongest of these rest-frame ultraviolet diagnostics in galaxies out to 
extremely high redshift.  
For example, the \ciii\ emission line doublet and the \civ\ doublet 
are redshifted into the NIRSpec range for redshifts of $z \ga 2.2$ and $z \ga 3.0$, respectively.  
With sufficiently deep integrations, \textit{JWST} should capture these and other 
ultraviolet diagnostics out to redshifts as high as galaxies can be found.  
Templates, like the composite presented here, will be necessary to interpret these \textit{JWST}
spectra, especially at the highest redshifts, where only a few of the brightest emission lines will 
probably be obtained.  

\acknowledgments  Acknowledgments: 
We thank the staff of Las Campanas Observatories for their dedicated service, which 
made possible these observations.  We thank the telescope allocation committees of the Carnegie 
Observatories, The University of Chicago, The University of Michigan, and Harvard University, 
for granting observing time to this observing program over several years.  We thank A.~Shapley,
C.~Steidel, G.~Zhu, and M.~Pettini for making available electronic versions of their spectra.
RB was supported by NASA through Hubble Fellowship grant \#51354 awarded by the Space 
Telescope Science Institute, which is operated by the 
Association of Universities for Research in Astronomy, Inc., for NASA, under contract 
NAS 5-26555.
\revisedApJ{We thank the referee for thoughtful suggestions that improved the paper.}

\bibliographystyle{astroads}
\bibliography{papers}  

\clearpage 

\begin{deluxetable}{lrllll}
\tabletypesize{\small}
\tablecolumns{6}
\tablewidth{0pc}
\tablenum{1}
\tablecaption{Comparison of the \megasaura\ stacked spectrum to other templates. \label{tab:stacked}}
\tablehead{
\colhead{spectrum}  & \colhead{$N_{stacked}$}   & \colhead{z} &  
\colhead{$\lambda_{rest}$ range} & \colhead{$R$} & \colhead{$SNR_{peak}$}\\
\colhead{} & \colhead{} & \colhead{} & \colhead{(\AA)} & \colhead{} & \colhead{}}
\startdata
\megasaura\ MagE stack (this paper) & 14     & $1.68<z<3.6$    & \revisedApJ{900}---3000  &  $\sim 3300$ & 103 \\
Shapley \etal\ (2003) composite     & 811    & $2.4<z<3.5$     & 920---2000  &  $\sim 560$  & 114\tablenotemark{a}\\
Jones \etal\ (2012) composite       & 81     & $3.5<z<4.5$     & 1000---1800 &  $\sim 660$  & 30\\
Steidel \etal\ (2010) composite     & 89, 102& $2.3\pm 0.3$    & 1000---1600 &   800, 1330  & \nodata \\
Steidel \etal\ (2016) composite     & 30     & $2.4\pm 0.11$   & 1000---2200 &   1400       & 38\\
cB58,  Pettini \etal\ (2002)        & 1      & $2.7276$        & 1075---2500 &   5200       & 55\\
eBOSS, Zhu \etal\ (2015)            & $8620$ & $0.6<z<1.2$     & 2200---7500 &  $\sim 2000$ & 70\tablenotemark{b} \\  
Stack of Chisholm \etal\ (2016) (this paper) & 41  & 0.0--0.25 & 1150---1780 &  \revisedApJ{3300}\tablenotemark{c}  & 150\\ 
\enddata
\tablecomments{Columns: 1) source of template spectrum; 2) Number of galaxies that
went into the template; 3) redshift range; 4) rest-frame wavelength range; 
5) spectral resolving power, defined as $R \equiv \delta \lambda / \lambda $, 
where $\delta \lambda$ is the full width at half maximum; 
6) peak signal-to-noise ratio of the continuum, per resolution element.}
\tablenotetext{a}{SNR not given in the paper; we have estimated it as the square root of the sample size, 
times the quoted average SNR of the input spectra.}
\tablenotetext{b}{SNR at 2500--3000~\AA .}
\tablenotetext{c}{\revisedApJ{The spectral resolution of the COS spectra varies with galaxy morphology.  
As described in the text, for each input galaxy we measured the effective spectral resolution from
Milky Way absorption lines, then convolved with a Gaussian and rebinned, to 
to produce an effective spectral resolution for the COS stack of $R=3300$.}}
\end{deluxetable}

\begin{deluxetable}{rrrrr}
\tabletypesize{\scriptsize}
\tablecolumns{5}
\tablewidth{0pc}
\tablenum{2}
\tablecaption{Emission lines. \label{tab:emissionlines}}
\tablehead{
\colhead{Line} & \colhead{$\lambda_{vac}$ (\AA)} & \colhead{$W_r$ (\AA)} & \colhead{uncertainty $W_r$ (\AA)} &
\colhead{significance}}
\startdata

   C~III~977 &   977.0200 &  $>$-0.1461 &               .. &           .. \\
 He~II~1084 &  1084.9420 &  $>$-0.0758 &               .. &           .. \\
Ly~$\alpha$ &  1215.6700 &     -1.2654 &             0.04 &        50.65 \\
   O~I~1304 &  1304.8576 &  $>$-0.0352 &               .. &           .. \\
  O~I*~1306 &  1306.0286 &  $>$-0.0348 &               .. &           .. \\
 Si~II~1309 &  1309.2757 &     -0.2441 &             0.02 &        21.26 \\
 C~II~1335a &  1334.5770 &   $>$-0.031 &               .. &           .. \\
C~II*~1335b &  1335.6630 &  $>$-0.0349 &               .. &           .. \\
C~II*~1335c &  1335.7080 &   $>$-0.035 &               .. &           .. \\
 N~II]~1430 &  1430.4100 &  $>$-0.0306 &               .. &           .. \\
  N~II]1431 &  1430.9730 &  $>$-0.0306 &               .. &           .. \\
 N~IV]~1486 &  1486.5000 &     -0.0428 &             0.02 &         4.12 \\
Si~II*~1533 &  1533.4312 &     -0.0367 &             0.01 &         3.41 \\
He~II~1640 broad &  1640.4170  & $-0.98$ &           0.1  &         --\\ 
He~II~1640 narrow &  1640.4170  & $>-0.22$  &        ..   &         .. \\ 
O~III]~1660 &  1660.8090 &     -0.0634 &             0.01 &         6.37 \\
O~III]~1666 &  1666.1500 &     -0.1716 &             0.02 &        13.98 \\
N~III]~1750 &  1749.7000 &  $>$-0.0286 &               .. &           .. \\
Si~III]~1882 &  1882.7070 &     -0.0673 &             0.01 &         5.70 \\
Si~III]~1892 &  1892.0290 &  $>$-0.0357 &               .. &           .. \\
{[C~III]}~1906 &  1906.6800 &     -0.5214 &             0.02 &        38.37 \\
 C~III]~1908 &  1908.7300 &     -0.3956 &             0.02 &        30.26 \\
  N II] 2140 &  2139.6800 &     -0.0649 &             0.02 &         4.46 \\
{[O III]} 2320 &  2321.6640 &  $>$-0.0571 &               .. &           .. \\
 C II] 2323 &  2324.2140 &  $>$-0.0533 &               .. &           .. \\
C II] 2325c &  2326.1130 &     -0.4034 &             0.09 &        19.61 \\
C II] 2325d &  2327.6450 &     -0.2055 &             0.07 &        10.35 \\
  C II] 2328 &  2328.8380 &      -0.064 &             0.04 &         3.17 \\
Si II] 2335a &  2335.1230 &  $>$-0.0608 &               .. &           .. \\
Si II] 2335b &  2335.3210 &  $>$-0.0625 &               .. &           .. \\
 Fe II* 2365 &  2365.5520 &      -0.183 &             0.02 &         9.40 \\
Fe II* 2396a &  2396.1497 &  $>$-0.0578 &               .. &           .. \\
Fe II* 2396b &  2396.3559 &  $>$-0.0576 &               .. &           .. \\
{[O II]} 2470 &  2471.0270 &     -0.3584 &             0.02 &        18.77 \\
Fe II* 2599 &  2599.1465 &  $>$-0.0596 &               .. &           .. \\
Fe II* 2607 &  2607.8664 &  $>$-0.0659 &               .. &           .. \\
Fe II* 2612 &  2612.6542 &  $>$-0.0665 &               .. &           .. \\
Fe II* 2614 &  2614.6051 &  $>$-0.0575 &               .. &           .. \\
Fe II* 2618 &  2618.3991 &  $>$-0.0632 &               .. &           .. \\
Fe II* 2621 &  2621.1912 &  $>$-0.0627 &               .. &           .. \\
Fe II* 2622 &  2622.4518 &  $>$-0.0676 &               .. &           .. \\
Fe II* 2626 &  2626.4511 &  $>$-0.0656 &               .. &           .. \\
Fe II* 2629 &  2629.0777 &   $>$-0.068 &               .. &           .. \\
Fe II* 2631 &  2631.8321 &  $>$-0.0672 &               .. &           .. \\
Fe II* 2632 &  2632.1081 &  $>$-0.0655 &               .. &           .. \\
Mg II 2797b &  2798.7550 &      -0.367 &             0.04 &        14.58 \\
Mg II 2797d &  2803.5310 &     -0.3522 &             0.04 &        14.20 \\
He I 2945   &  2945.1030 &  $>$-0.1063 &               .. &           .. \\
\enddata
\tablecomments{Rest-frame ultraviolet emission lines in the \shapenorm\ stacked
spectrum.  Columns are: 1) line identification, 2) vacuum wavelength (\AA), 
3) measured rest-frame equivalent width (\AA) (with negative 
sign indicating absorption), 4) uncertainty on the previous column,
5) significance of the detected feature, in $\sigma$, 
according to the \citet{Schneider:1993hx} significance critereon.
\revisedApJ{Non-detections are quoted as $3\sigma$ limits.}
Note: The C~II~232X complex is challenging to fit given the close 
spacing of the four emission lines. 
The total equivalent width of the complex is well-measured, at 
$W_r = -0.68 \pm 0.14$, but the fit is degenerate as to which components
contain most of the flux.  
We therefore fit the complex by fixing the relative strengths of the lines to 
the predictions of MAPPINGS~V. 
}
\end{deluxetable}
  
%
%
\begin{deluxetable}{llllllllllll}
\tabletypesize{\scriptsize}
\tablecolumns{10}
\tablenum{3}
\tablecaption{Absorption velocity measurements. \label{tab:vmax}}
\tablehead{
\colhead{transition}       & \colhead{IP} & \colhead{$v_{mean}$} & \colhead{$\sigma$} & \colhead{$v_{max}$} & \colhead{$\sigma$} & \colhead{$v_{mean}$} & \colhead{$\sigma$} & \colhead{$v_{max}$} & \colhead{$\sigma$}\\
       & \colhead{(eV)} & \colhead{(\kms)}  & \colhead{(\kms)}  & \colhead{(\kms)}  & \colhead{(\kms)} & \colhead{(\kms)}  & \colhead{(\kms)}  & \colhead{(\kms)}  & \colhead{(\kms)}}
\startdata
\cutinhead{\megasaura\ MagE/Magellan shape-normalized stack} 
& & \multicolumn{4}{l}{Measurements of weighted average stack} & \multicolumn{4}{l}{Measurements of median stack}\\
O I 1302    &  13.6  & \nodata  & \nodata & $<-500$\tablenotemark{a} & \nodata & \nodata &   \nodata & $<-500$\tablenotemark{a} & \nodata \\
Mg II 2796  &  15.0    &  $-310$ &        3 &  $-885$ &       0 &    $-271$ &       10 &  $-721$ &      72 \\   
Fe II 2344  &  16.2    &  $-225$ &       20 &  $-965$ &     111 &    $-178$ &       15 &  $-741$ &      50 \\  
Fe II 2383  &  16.2  & \nodata  & \nodata & $<-840$\tablenotemark{b} & \nodata & \nodata &    \nodata & $<840$\tablenotemark{b} & \nodata\\ 
Si II 1260  &  16.35   &   $-97$ &        1 &  $-772$ &      37 &    $-203$ &        1 &  $-740$ &      24  \\  
Si II 1526  &  16.35   &   $-92$ &        1 &  $-632$ &      30 &    $-138$ &        4 &  $-606$ &      39  \\ 
Al II 1670  &  18.8    &  $-128$ &       18 &  $-707$ &      40 &    $-145$ &       34 &  $-634$ &      27  \\ 
C II 1334   &  24.4    &   $-53$ &       13 &  $-938$ &     183 &    $-163$ &       25 &  $-971$ &     227  \\ 
Al III 1854 &  28.4    &  $-568$ &      136 & $-2268$ &     494 &    $-275$ &      148 &  $-984$ &     513  \\ 
Si IV 1393  &  45.1    &  $-260$ &       76 & $-1327$ &     406 &    $-419$ &       67 & $-2097$ &     290  \\ 
C IV 1548   &  64.49   &  $-517$ &       43 & $-2696$ &     159 &    $-535$ &       41 & $-2513$ &     150  \\ 
\cutinhead{COS/\textit{HST} $R=3300$ shape-normalized stack}\\
& & \multicolumn{4}{l}{Measurements of weighted average stack} & \multicolumn{4}{l}{Measurements of median stack}\\
O I 1302    &  13.6  &  \nodata  &  \nodata & $<-500$\tablenotemark{a} & \nodata &   \nodata & \nodata & $<-500$  & \nodata\\  
Si II 1260  &  16.35 & $-189$ &        6.0 &  $-815$ &      20 &                     $-179$ &        6 &  $-906$ &      48 \\
Si II 1526  &  16.35 & $-107$ &        3.0 &  $-574$ &       0 &                     $-114$ &        0 &  $-532$ &       0 \\
Al II 1670  &  18.8  &  $-64$ &       34.0 &  $-431$ &      51 &                     $-88$ &       61 &  $-399$ &       0 \\
C II 1334   &  24.4  & $-191$ &       17.0 & $-1561$ &      86 &                     $-104$ &        2 &  $-684$ &      34 \\
Si IV 1393  &  45.1  & $-345$ &       49.0 & $-1833$ &     329 &                     $-371$ &       10 & $-1733$ &      51 \\
C IV 1548   &  64.5  & $-355$ &       27.0 & $-2386$ &     172 &                     $-516$ &       28 & $-2441$ &     154 \\
 \enddata
\tablecomments{Columns are: 1) Line label; 2) ionization potential in electron volts; 
3) absorption-weighted mean velocity \vmean\ for the weighted average stack;  4) uncertainty in \vmean ;
5) maximum velocity \vmax\ for the weighted average stack;  6) uncertainty in \vmax ;
7)--10) same as 3)--6) but for the median stack.  
N~V~1238 is not listed because absorption was not clearly detected. 
\tablenotetext{a}{Blend with photospheric absorption.}
\tablenotetext{b}{Blend with Fe~II~2374 absorption.}
}
\end{deluxetable}

\clearpage 

\clearpage 



\clearpage
\begin{subfigures}
\begin{figure}
\includegraphics[page={1}, scale=0.6]{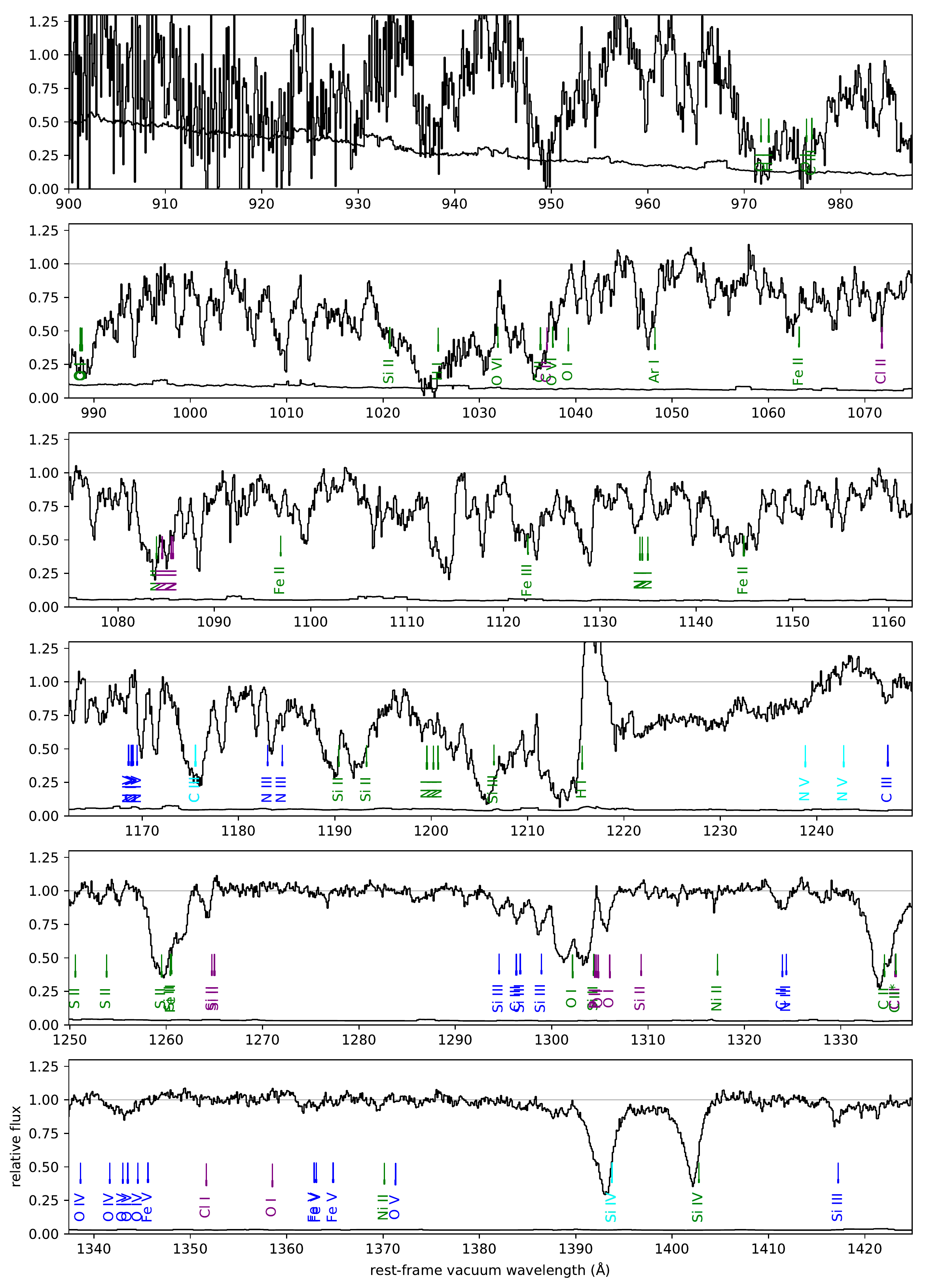} 
\end{figure}
\begin{figure}
\includegraphics[page={2}, scale=0.6]{f3.pdf} 
\end{figure}
\begin{figure}
\includegraphics[page={3}, scale=0.6]{f3.pdf} 
\end{figure}
\begin{figure}
\includegraphics[page={4}, scale=0.6]{f3.pdf} 
\caption{The \megasaura\ \shapenorm\ stacked spectrum.  
The systemic redshift for most of the input galaxies was determined from the 
\ciii\ emission lines. 
Individual spectra were normalized by the hand-fit continuum before stacking; 
the continuum level is shown as the grey horizontal line.
Lines of interest are marked at the systemic redshift, and color-coded as follows: 
interstellar medium (ISM) \textit{(green)}; photospheric absorption \textit{(blue)};
stellar wind lines \textit{(cyan)}; nebular emission \textit{(red)};  
and fine structure emission \textit{(purple)}. 
\label{fig:multipanel_spectrum}}
\end{figure}
\end{subfigures}

\clearpage 
\begin{figure}
\includegraphics[width=3.3in]{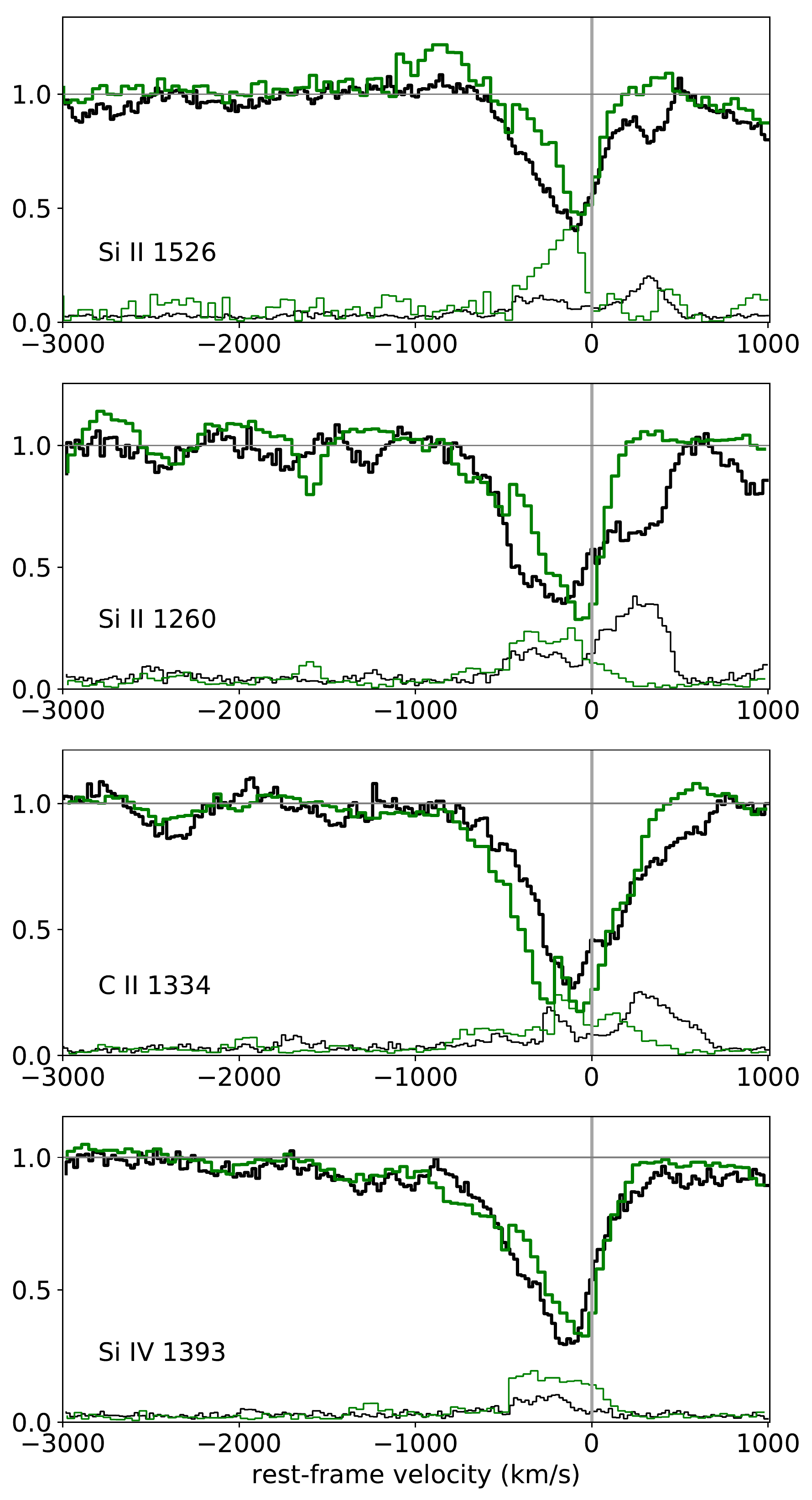} 
\includegraphics[width=3.3in]{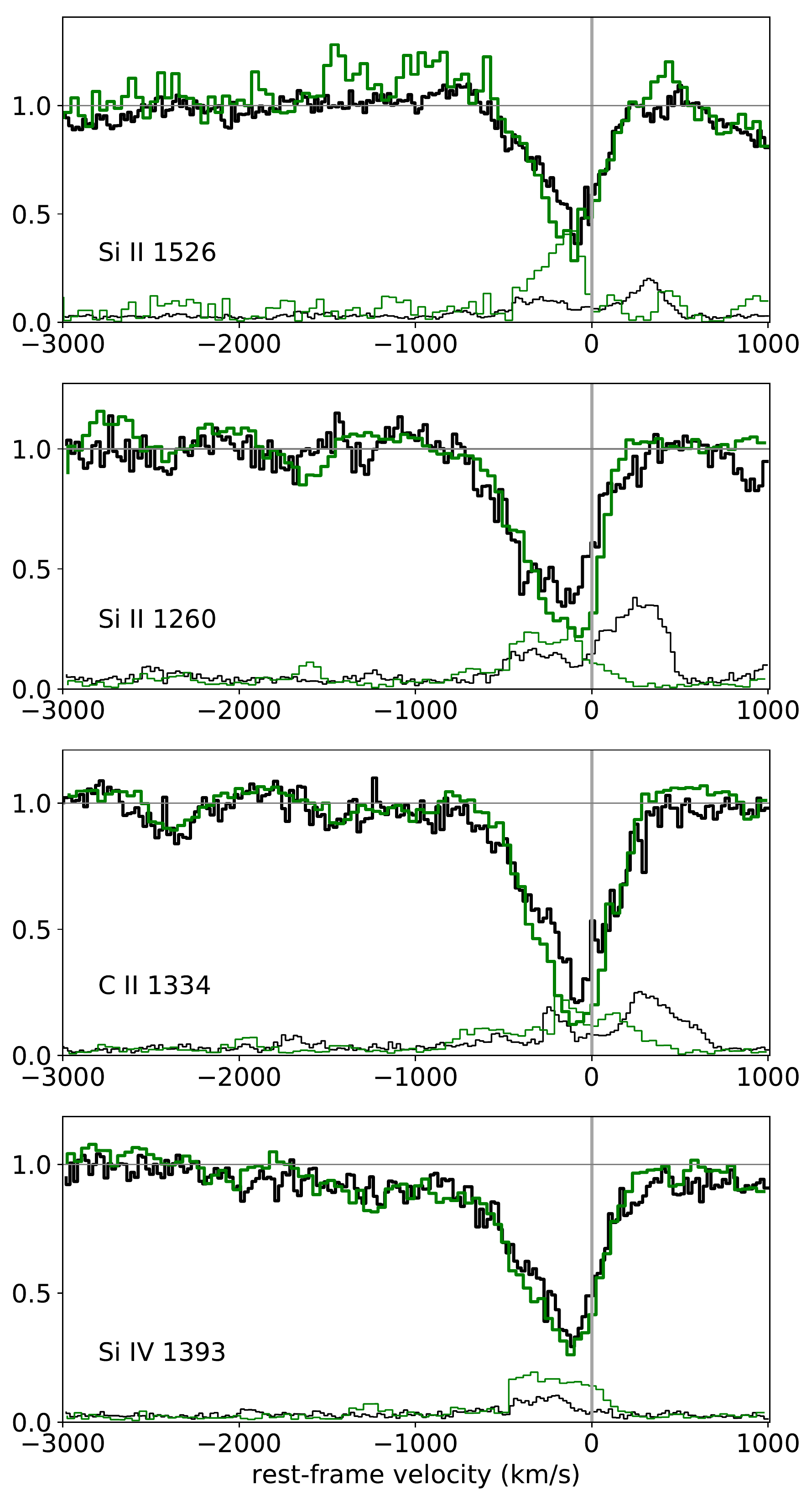} 
\figcaption{Comparison of the velocity profiles of strong absorption features at two 
cosmic epochs:  at $z\sim2$ from the \shapenorm\ weighted-average \megasaura\ stack \textit{(black steps)}, and 
our stack of $z\sim0$ \textit{HST}/COS spectra \textit{(green steps)} from \citet{Chisholm:2016bi}.  
The left panel shows the weighted average for each stack; the right panel shows the median.
In all cases, we plot the jackknife uncertainty as the uncertainty spectrum. 
Excess variance is seen at the locations of  Si II* 1264  and C~ II*~1335.  
The velocity profiles of the $z\sim 2$ stack are very similar to that of the $z\sim 0$ stack.
The input spectra were continuum--normalized before stacking, so unity represents zero absorption. 
\label{fig:likeheckmanfig}}
\end{figure}

\begin{subfigures}
\begin{figure}
\includegraphics[width=3.5in]{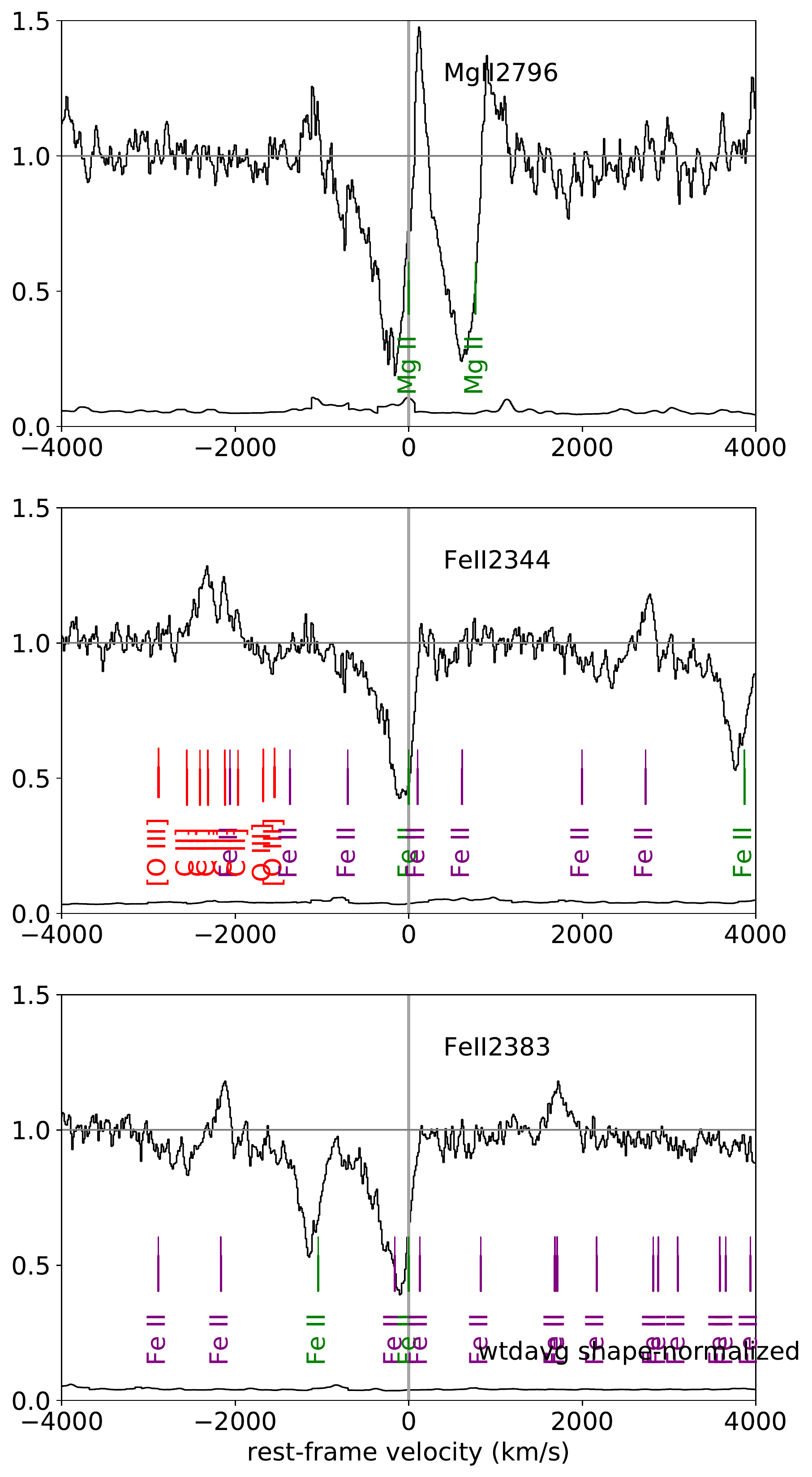} 
\includegraphics[width=3.5in]{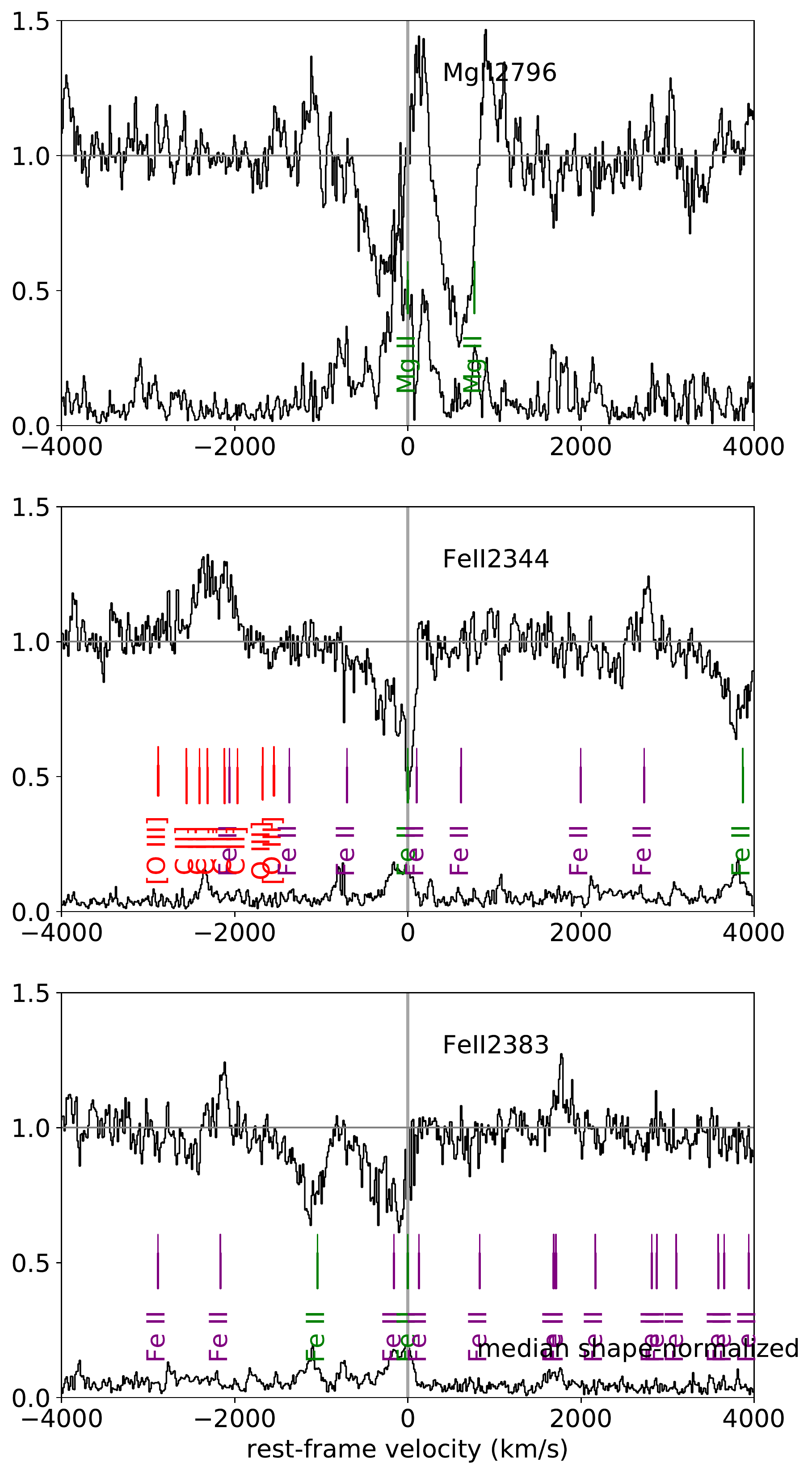} 
\caption{The \megasaura\ \shapenorm\ stacked spectrum, for a weighted average (left panel), and a median (right panel) stack. 
Plotted is $f_{\nu}$ versus rest-frame velocity, with zero velocity at the systemic redshift.  
Since the spectra were continuum--normalized before stacking, unity represents zero absorption. 
Overplotted are two characterizations of the uncertainty:
the propagation of the individual uncertainty spectra (left panel), 
and the jackknife uncertainty (right panel). 
Vertical ticks and labels mark the expected positions of other lines of interest, color-coded as 
in Figure~\ref{fig:multipanel_spectrum}.  The plotted transitions trace large-scale galactic winds,  and have 
ionization potentials of 15.0 eV (Mg~II) and 16.2~eV (Fe~II). 
\label{fig:windsb}}
\end{figure}
\begin{figure}
\includegraphics[width=3.5in]{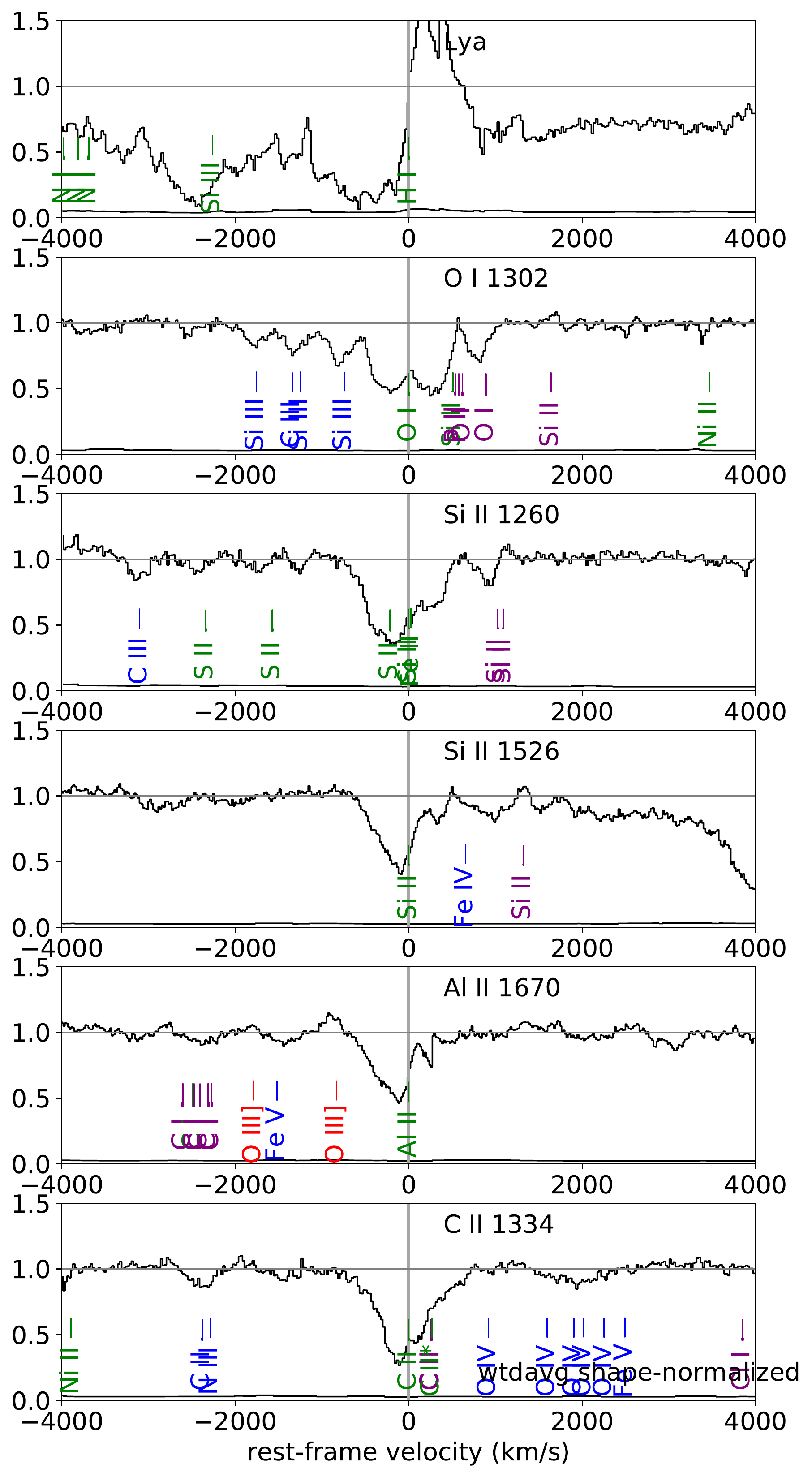} 
\includegraphics[width=3.5in]{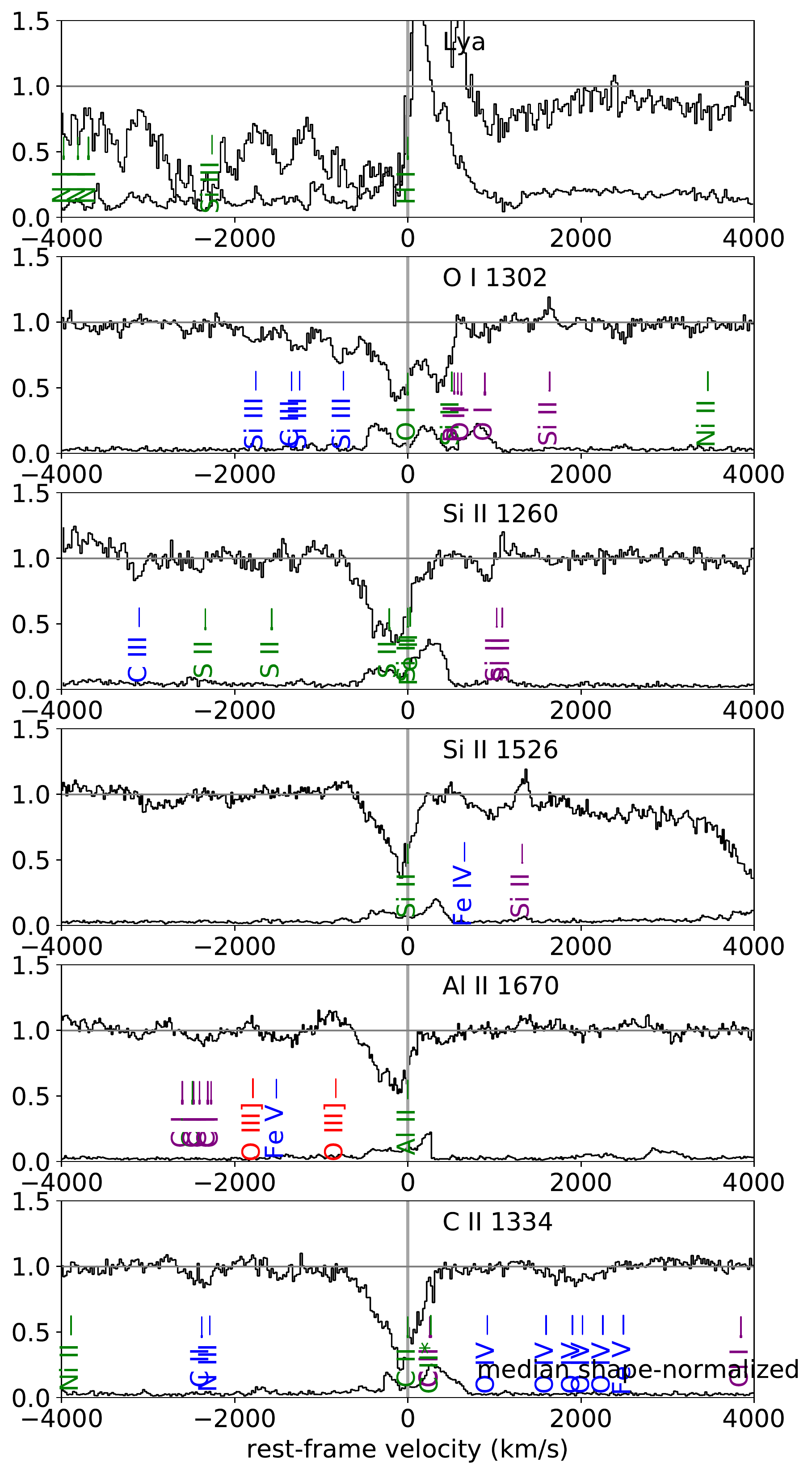} 
\caption{Same as Figure~\ref{fig:windsb}, for transitions with ionization potentials from 13.6~eV (top panel) 
to 24.4~eV (bottom panel).  The weighted average stack is shown in the left panel, and the median stack in the right panel.
\label{fig:windsa}}
\end{figure}
\begin{figure}
\includegraphics[width=3.5in]{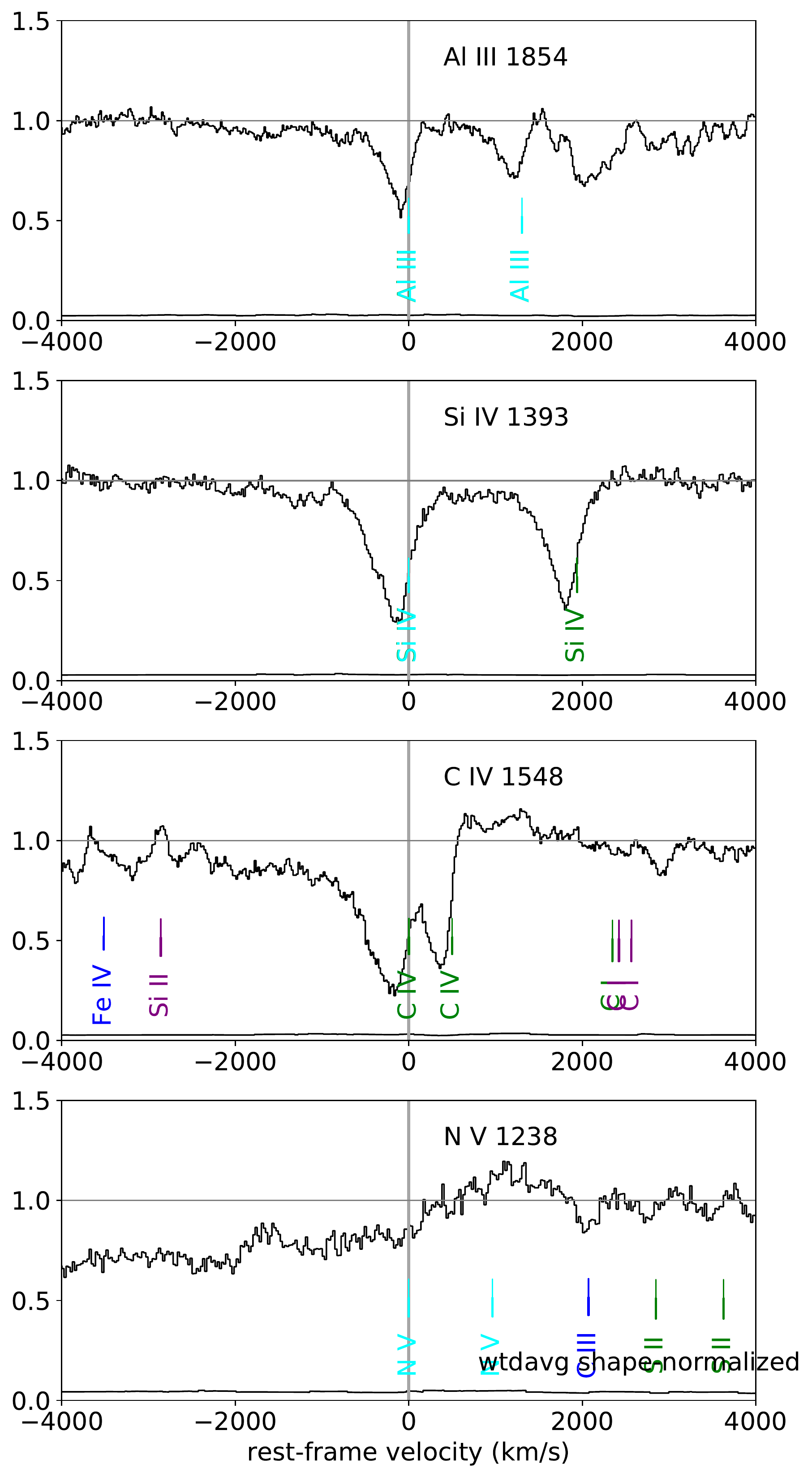} 
\includegraphics[width=3.5in]{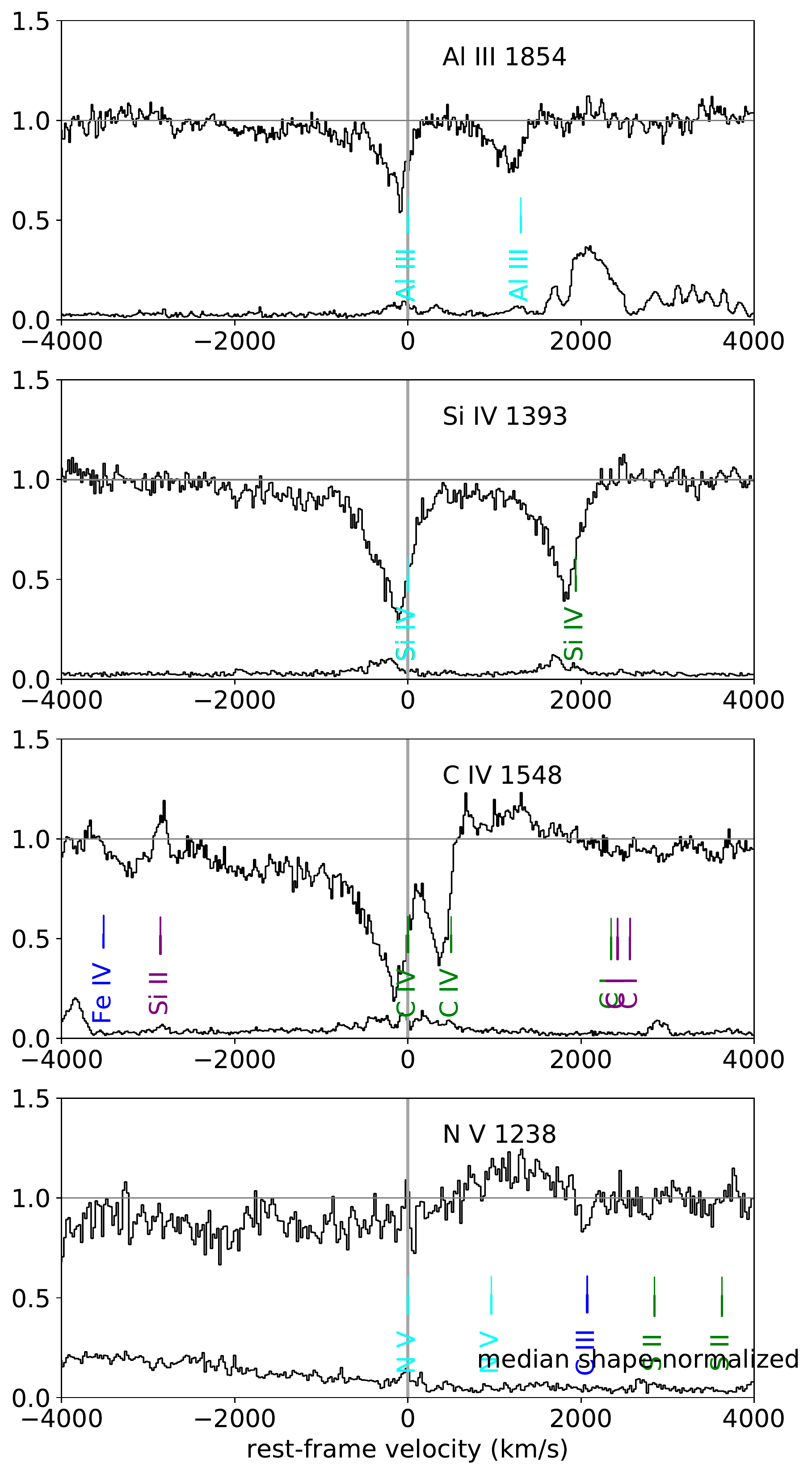} 
\caption{Same as Figure~\ref{fig:windsb}, for transitions with high ionization potential.  
Transitions are plotted in order of increasing ionization potential, from top to bottom, from 
28.4 to 97.9~eV.  
 The weighted average stack is shown in the left panel, and the median stack in the right panel.
\label{fig:windsc}}
\end{figure}
\end{subfigures}

\begin{figure}
\includegraphics[width=5in]{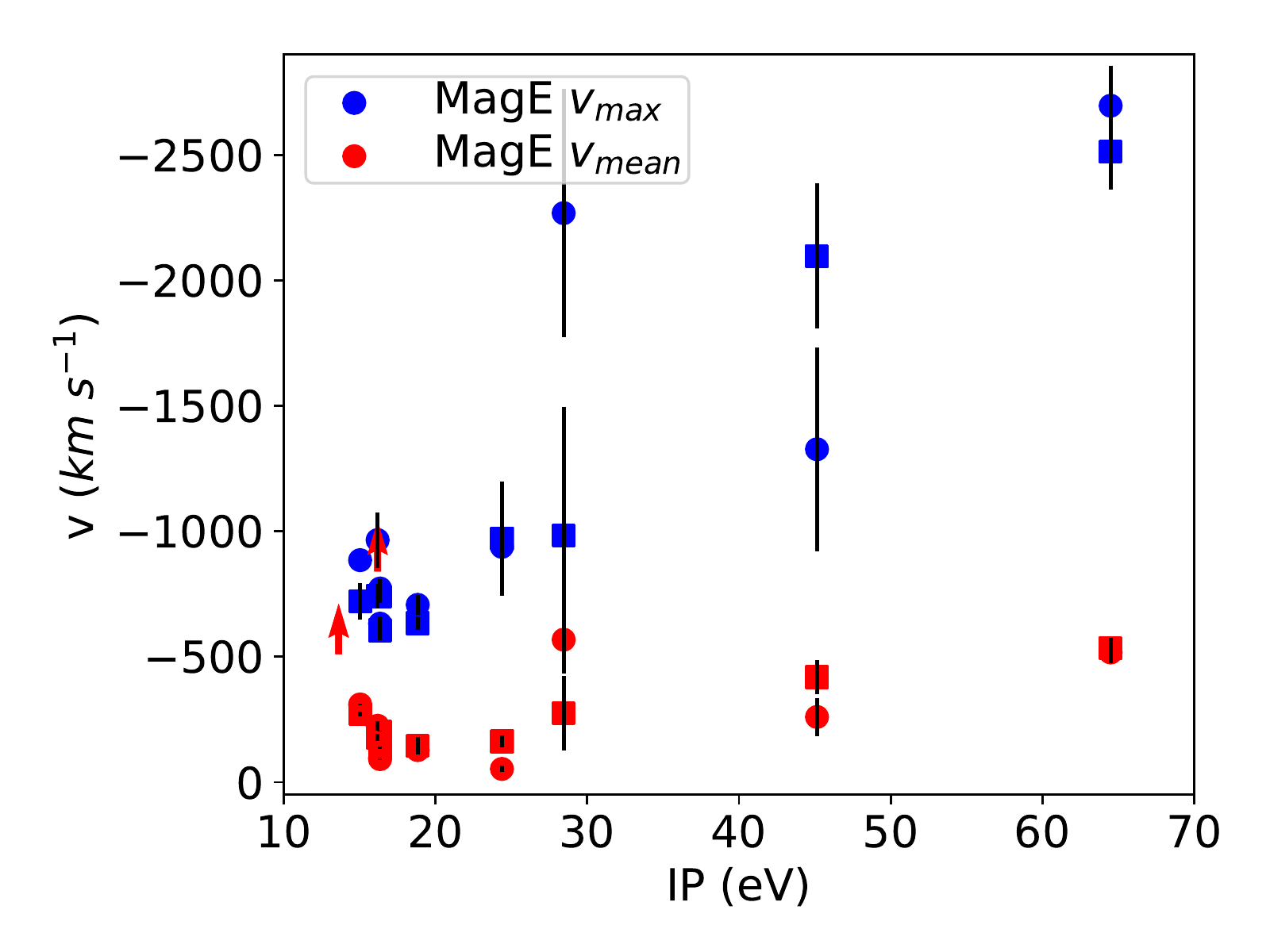} 

\includegraphics[width=5in]{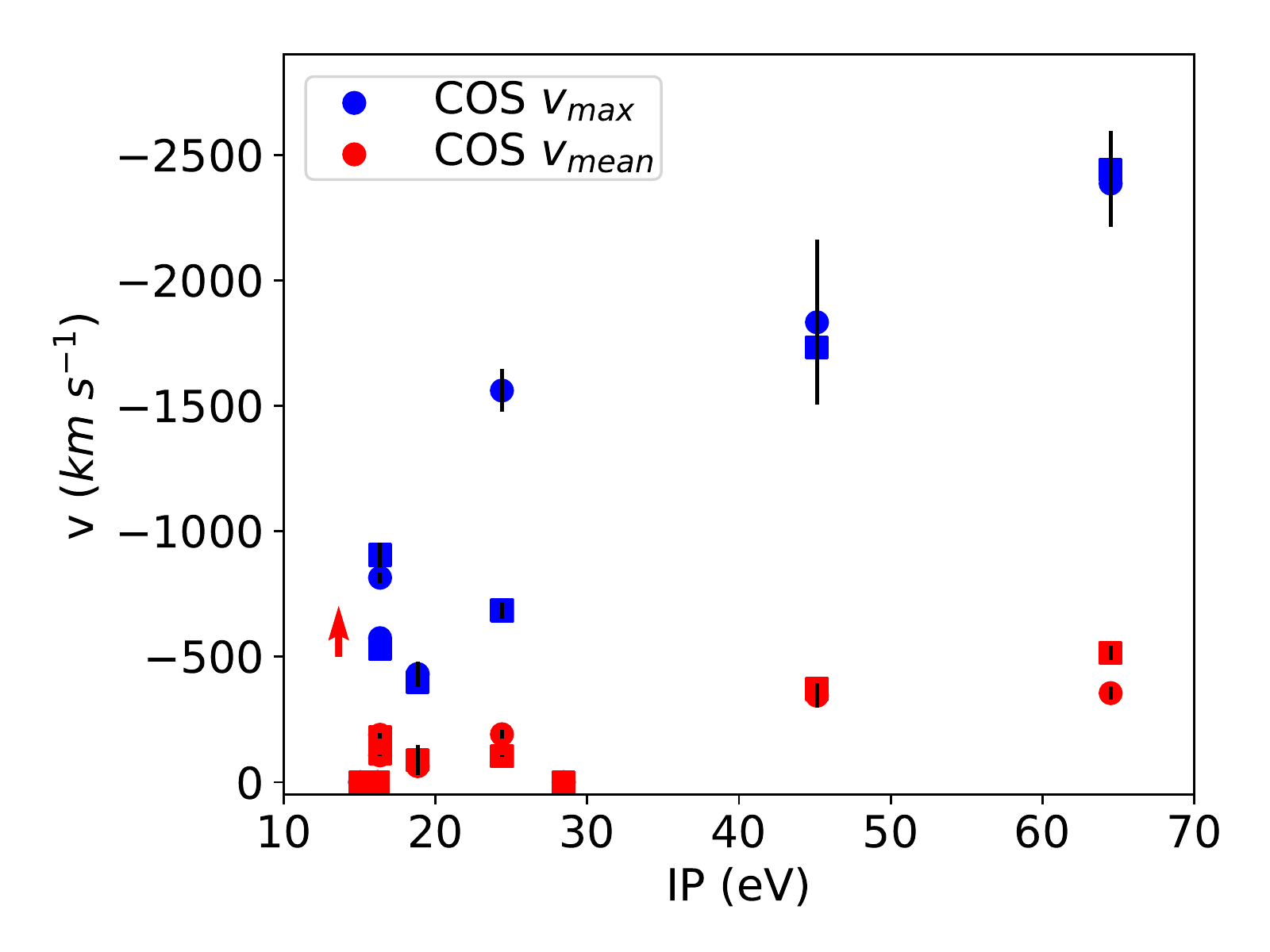} 
\figcaption{Outflow velocity versus ionization potential.  Measurements of 
the \megasaura\ MagE shape-normalized stacks ($z\sim2$) are plotted in the
top panel; measurements of the \coshst\ stack ($z\sim0$) are plotted in the lower panel.
In each panels, circles show measurements for the weighted-average stack, and squares 
show measurements to the median stack.  
Colors differentiate two measures of the outflow velocity:
the absorption-weighted mean velocity \vmean\ \textit{(red symbols)} and the maximum velocity \vmax\ \textit{(blue symbols)}.
\label{fig:vIP}}
\end{figure}

\begin{subfigures}
\begin{figure}
\includegraphics[width=3.5in]{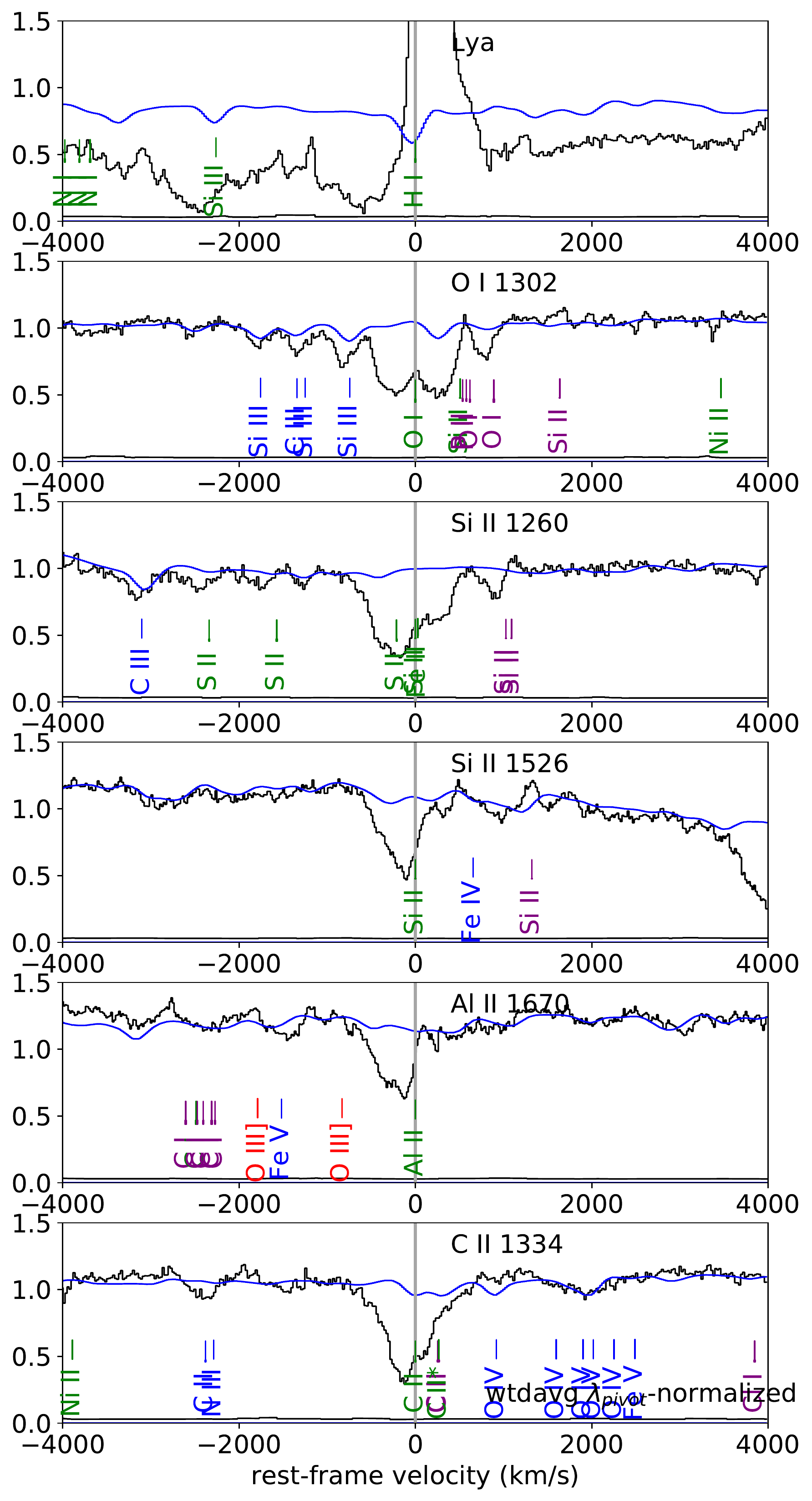} 
\includegraphics[width=3.5in]{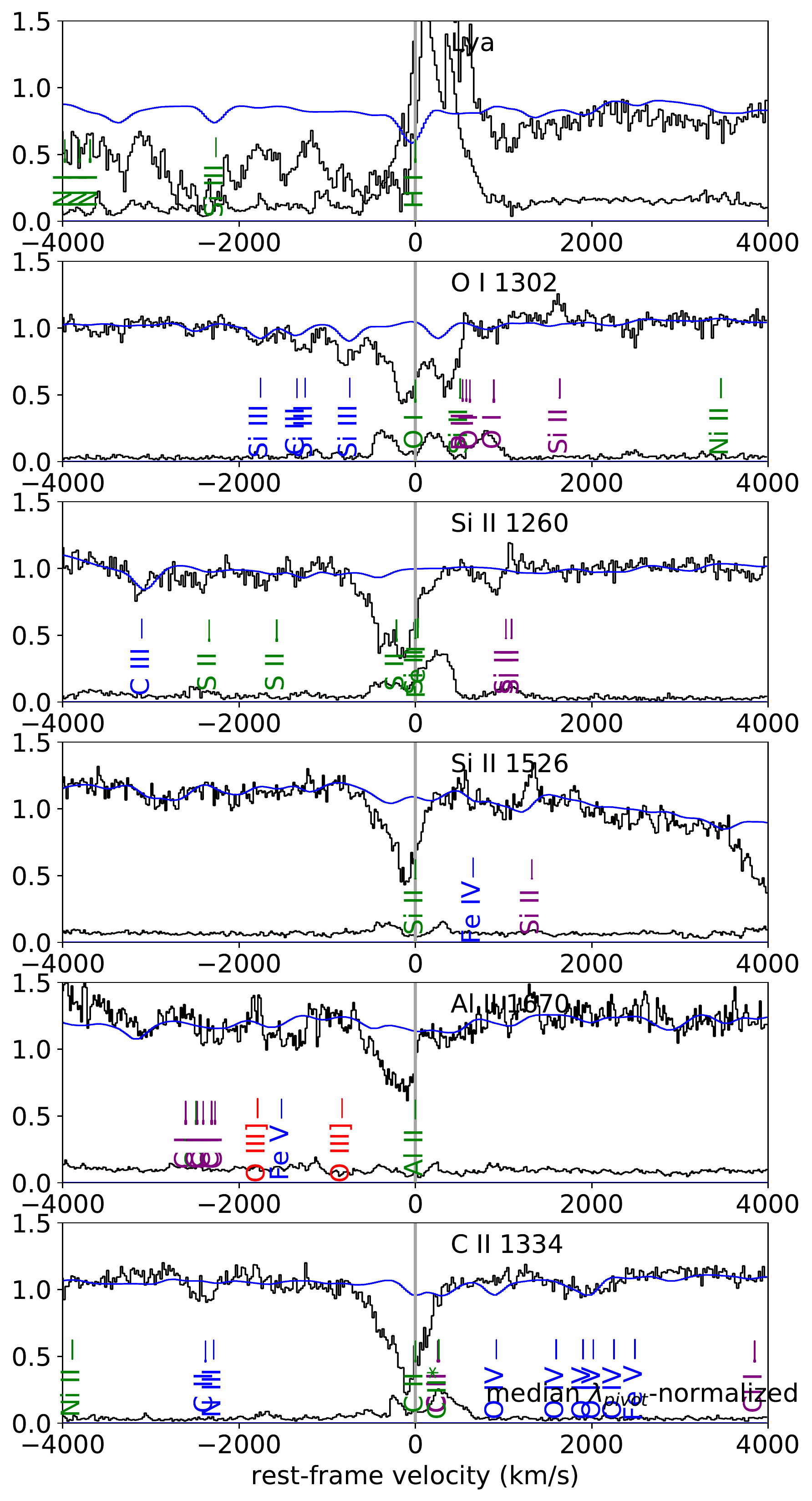} 
\caption{The \megasaura\ \pivot\ stacked spectrum  (in black), compared to the best-fit Starburst99 stellar
continuum (blue).  Transitions plotted are as in Figure~\ref{fig:windsb}, for the weighted average (left panel), and median (right panel) stacks.
\label{fig:s99a}}
\end{figure}
\begin{figure}
\includegraphics[width=3.5in]{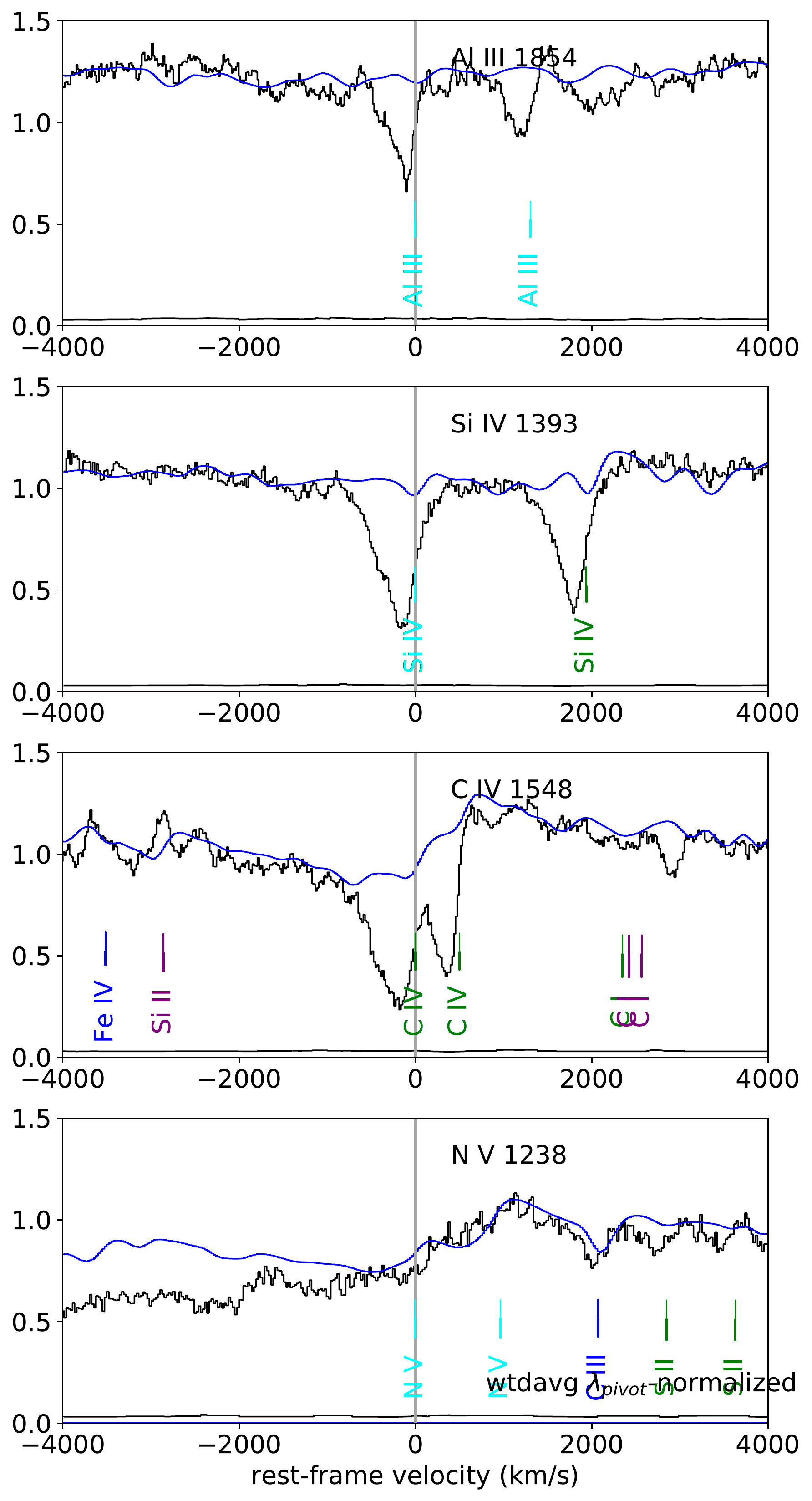} 
\includegraphics[width=3.5in]{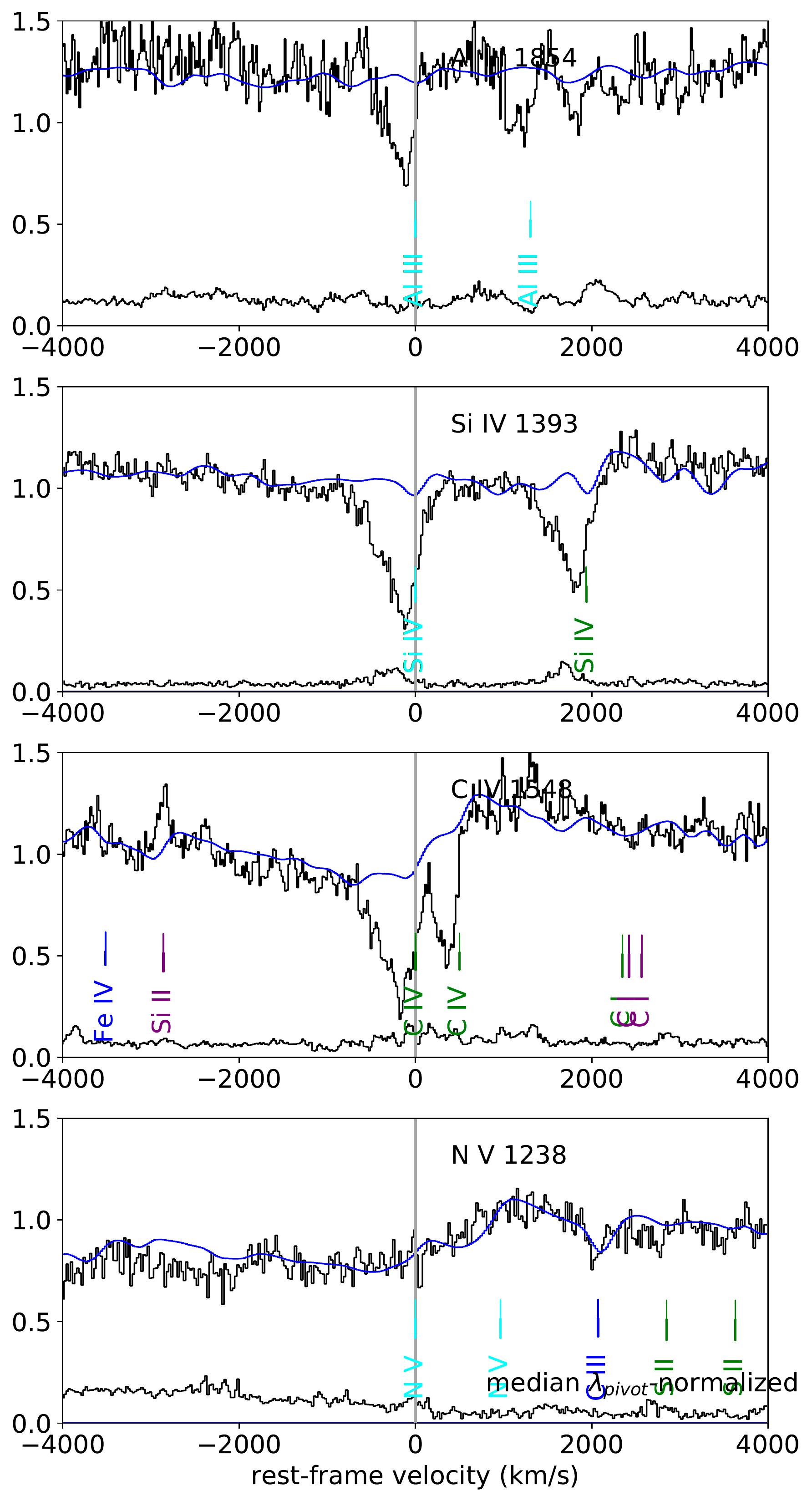} 
\caption{Same as Figure~\ref{fig:s99a}, for high ionization potential transitions,  
for the weighted average (left panel), and median (right panel) stacks.
 The S99 fit reproduces the blue wing of the C~IV absorption fairly well;  
it does a poorer job reproducing the N~V feature and the blue wing of the Si~IV feature.  
The poor fit blueward of N~V~1238 is due to broad Lyman $\alpha$ absorption and H$_2$ extinction.   
\label{fig:s99c}}
\end{figure}
\end{subfigures}

\begin{figure}
\includegraphics[width=3.5in]{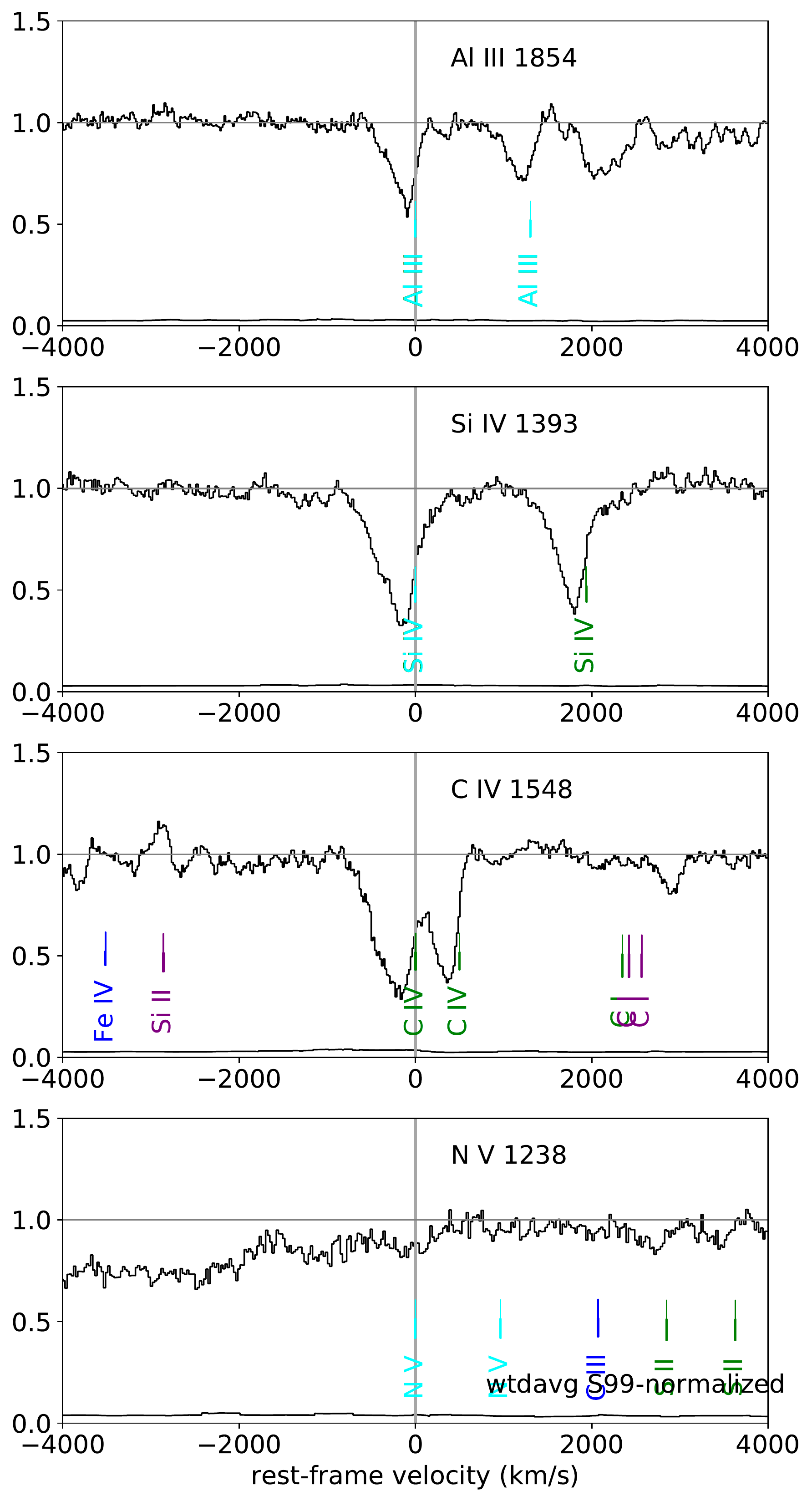} 
\includegraphics[width=3.5in]{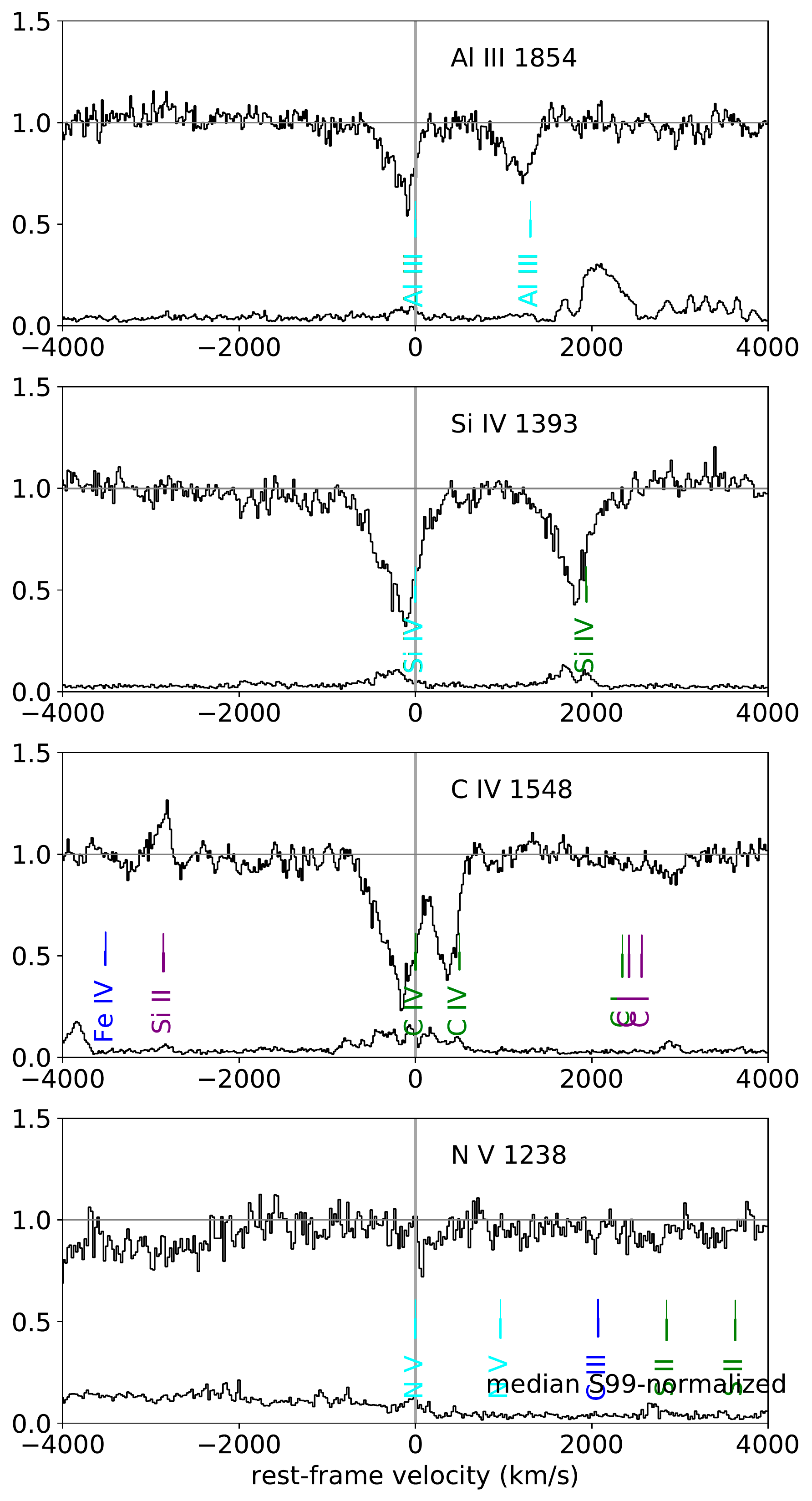} 
\figcaption{The  \snine\ stack, for transitions with high ionization potential, for the
weighted average (left panel) and median (right panel) stacks.
The profiles show negligible absorption at high velocities, which confirms that the best-fit Starburst99 models are able 
reproduce the stellar winds.
\label{fig:snine}}
\end{figure}

\end{document}